\documentclass[twoside,11pt]{article}

\usepackage[accepted]{melba}

\usepackage{amsmath,amsfonts}
\usepackage{amsmath,amsfonts, booktabs}
\usepackage{pdflscape}
\usepackage{hyperref}
\usepackage[all]{hypcap}
\usepackage[normalem]{ulem}
\melbaid{2024:013}  % This is provided upon by the publishing editor
\doi{https://doi.org/10.59275/j.melba.2024-151b}
\melbaauthors{Dimitrijevic, Noblet and De Leener}  % Note: this one is also used to set the pdf 'authors' metadata
\volume{2}
\firstpageno{916}  % Communicated by the publishing editor
\melbayear{2024}  % The publication year
\datesubmitted{09/2023}  % Date submitted to MELBA: mm/yyyy
\datepublished{06/2024}  % Today's date: mm/yyyy

\melbaspecialissue{Special Issue on Image Registration}
\melbaspecialissueeditors{Veronika Zimmer, Alessa Hering, Mattias Heinrich}

\ShortHeadings{Impact of Initialization on Pediatric Brain Intra-subject MR Registration}{Dimitrijevic, Noblet and De Leener}

\title{\rev{Impact of Initialization on Intra-Subject Pediatric Brain MR Image Registration: A Comparative Analysis between \\ SyN ANTs and Deep Learning-Based Approaches}}

\author{\firstname Andjela \surname Dimitrijevic\orcid{0000-0002-8799-6300} \email andjela.dimitrijevic@polymtl.ca \\  % start right after \author{, or there will be an extra space
	\addr NeuroPoly Lab, Institute of Biomedical Engineering, Polytechnique Montréal, Montréal, QC, Canada \\
        \addr Research Center, Ste-Justine Hospital University Centre, Montréal, QC, Canada
	\AND
	\name Vincent Noblet\orcid{0000-0002-3655-3163} \email vincent.noblet@unistra.fr \\
	\addr ICube-UMR 7357, Université de Strasbourg, CNRS, Strasbourg, France
        \AND
	\name Benjamin De Leener\orcid{0000-0002-1378-2756} \email benjamin.de-leener@polymtl.ca \\
	\addr NeuroPoly Lab, Institute of Biomedical Engineering, Polytechnique Montréal, Montréal, QC, Canada \\
        \addr Research Center, Ste-Justine Hospital University Centre, Montréal, QC, Canada \\
        \addr Computer Engineering and Software Engineering, Polytechnique Montréal, Montréal, QC, Canada
}

\begin{document}

% top matter
\maketitle

% abstract
\begin{abstract}%   <- trailing '%' for backward compatibility of .sty file
This study evaluates the performance of \finalrev{conventional SyN ANTs and learning-based} registration methods in the context of pediatric neuroimaging, specifically focusing on intra-subject deformable registration. The comparison involves three approaches—without (NR), with rigid (RR), and with rigid and affine (RAR) initializations. \finalrev{In addition to initialization, performances are evaluated in terms of accuracy, speed, and the impact of age intervals and sex per pair. Data consists of the} publicly available \rev{MRI scans from the} Calgary Preschool dataset, which includes 63 children aged 2-7 years, allowing for 431 registration pairs. We implemented the unsupervised \finalrev{deep learning (DL) framework with a }U-Net \finalrev{architecture} using DeepReg and it was 5-fold cross-validated. The evaluation includes Dice scores for tissue segmentation from 18 smaller regions obtained by SynthSeg, analysis of log Jacobian determinants, and registration \rev{pro-rated training and inference times}. \rev{Learning-based approaches, with or without linear initializations, exhibit slight superiority over SyN ANTs in terms of Dice scores. Specifically, DL-based implementations with RR and RAR initializations \finalrev{significantly} outperform SyN ANTs.} \finalrev{The lower Dice scores of SyN ANTs are likely due to its lack of population-based optimization, unlike the DL methods which learn optimal parameters through training. Both SyN ANTs and DL-based registration involve parameter optimization, but the choice between these methods depends on the scale of registration—network-based for broader coverage or SyN ANTs for specific structures. Learning-based registration offers fast inference times but needs training, whereas SyN ANTs requires manual fine-tuning, with less clear guidelines, particularly for younger cohorts. Both methods face challenges with larger age intervals due to greater growth changes. Future work will extend the framework to younger populations and explore models that better separate different levels of transformations for improved local brain region registration. The main takeaway is that while DL-based methods show promise with faster and more accurate registrations, SyN ANTs remains robust and generalizable without the need for extensive training, highlighting the importance of method selection based on specific registration needs in the pediatric context.}
Our code is available at~\url{https://github.com/neuropoly/pediatric-DL-registration}. 
% \sout{with mean Dice scores of 0.825±0.024 versus 0.818±0.017 and 0.913±0.018 versus 0.895±0.023, respectively.}
\end{abstract}
% The observed lower Dice scores may stem from the non-optimization of ANTs in a population-based manner, contrasting with the network-based registration's reliance on learning optimal parameters through training. Both SyN ANTs and DL-based registration involve parameter optimization, but the choice between these methods depends on the scale of registration—network-based for broader coverage or SyN ANTs for specific structures. Learning-based registration proves fast at inference but necessitates training, while SyN ANTs requires manual fine-tuning with unclear guidelines, especially for younger cohorts. As the age interval increases, both methods show difficulties in registering greater growth changes.} Future directions include extending the framework to younger populations and exploring models that can separate different levels of transformations for improved registration of local brain regions.
	%

% keywords

% Introduction (or first section)
\section{Introduction}
	Deformable image registration involves the alignment of a pair of images to establish a shared coordinate reference framework. It is used for both intra and inter-subject analyses within the medical domain, playing a vital role in achieving normalized visualizations across brain scans~\citep{Uchida-2013}. This research focuses on the fact that so far, deformable image registration is less adapted to pediatric data, this can arise from bigger volume differences when analyzing longitudinal data of a subject’s brain at two different time-points, but also the lack of pediatric data availability~\citep{Barkovich2019-bm}. However, improving registration performed on neuroimaging data of pediatric populations remains essential for template creation as well as different diagnostic pipelines. Currently, conventional deformable registration methods such as ANTs~\citep{Avants2011-xu}, NiftyReg~\citep{Modat2010-qn} or Elastix~\citep{Klein2010-hx} are functional. Nonetheless, when dealing with extensive datasets, the iterative optimization-based estimation of deformation fields makes the process time-intensive. \rev{The emerging deep learning (DL)-based techniques,} incorporating convolutional neural networks (CNN)\rev{,} can allow faster registrations by applying a learning-based approach instead. In essence, these recently devised techniques enable the direct estimation of deformation fields from input 3D \revision{volume} pairs. \finalrev{This study aims to evaluate registration implementations in the pediatric context, comparing the conventional SyN ANTs method with DL-based approaches, with a focus on their performance in terms of accuracy, speed, initialization, and the impact of age intervals per pair and separated by sex within intra-subject pediatric data.}

\subsection{Current DL-based Registration Frameworks for Deformable Registration \finalrev{and Its Evaluation in the Pediatric Context}}
\finalrev{A popular DL-based registration approach is VoxelMorph~\citep{Balakrishnan2019-sr}, which is designed for brain MRI applications. It uses a U-Net-like architecture (encoder-decoder with skip-connections)\citep{Ronneberger2015-wd} and employs the scaling and squaring integration method on computed velocity fields to obtain diffeomorphic deformation fields. Kuang et al.\citep{Kuang2019-ru} developed the fast image registration (FAIM) algorithm, which showed superior results to VoxelMorph. FAIM, composed of a spatial deformation module largely inspired by the spatial transformer networks (STN)~\citep{Jaderberg2015-fz}, stacks moving and fixed images as input for the network and uses a training loss composed of a cross-correlation metric and regularization to ensure smooth, non-negative Jacobian determinants. Similarly, \cite{Zhang2018-lq} introduced the inverse-consistent deep network (ICNet) with inverse-consistent and anti-folding constraints added to a mean squared distance similarity metric. Also using a U-Net like architecture and STN, their method outperformed Demons-based registration~\citep{THIRION1998243} as well as \revision{symmetric normalization (SyN) based registration~\citep{Avants2008}}.

All these DL-based registration frameworks often use U-Net-like architectures which frequently outperform conventional registration methods and are trained on adult MRI brain data. Now, these tools are even incorporated into popular neuroimaging analysis tools such as FreeSurfer. EasyReg~\citep{Iglesias2023-ng} implements SynthSeg's~\citep{Billot2023-po} model to predict forward and backward nonlinear fields, achieving symmetric and diffeomorphic transformations. This approach is the first learning-based registration algorithm publicly available via FreeSurfer~\citep{Fischl2012-qx}.
% These U-Net-like architectures are often used for their strong performance in the mentioned DL-based registration frameworks and frequently outperform conventional registration methods, such as Demons-based registration~\citep{THIRION1998243} and symmetric normalization (SyN) based registration~\citep{Avants2008}. For instance, EasyReg~\citep{Iglesias2023-ng} implements SynthSeg's~\citep{Billot2023-po} model to predict forward and backward nonlinear fields, achieving symmetric and diffeomorphic transformations. This approach is the first learning-based registration algorithm publicly available via FreeSurfer~\citep{Fischl2012-qx}.

However, few of these registration implementations have been evaluated in the pediatric longitudinal context~\citep{Ghosh2010-si}. \rev{Efficient and accurate registration algorithms could facilitate the comprehensive analysis of longitudinal changes in deformation fields across a larger number of subjects, particularly when handling extensive neuroimaging datasets.} This gap presents an opportunity to analyze neurodevelopment in young children, where these advanced DL-based registration methods could be particularly impactful. Recently, \cite{Hoffmann2023-jf} use SynthMorph to evaluate affine and joint registration (affine+deformable) \finalrev{across six different datasets, including only two that encompass pediatric subjects aged 5-21 years from the Lifespan Human Connectome Project Development (HCPD) and MASiVar (MASi) datasets (total of 100 pediatric images). Notably, these pediatric datasets were not used for training, which was conducted solely with adult populations. Moreover, the pediatric datasets did not undergo any preprocessing steps such as skull-stripping or N4 correction, which could impact the registration results. Despite these limitations, SynthMorph demonstrated successful generalization, achieving Dice scores between 0.85 and 0.90 on 23 bilateral brain regions within these pediatric datasets. Nevertheless, further evaluation of learning-based methods contrasted to conventional ones is needed, as significant neurodevelopmental changes can be revealed in longitudinal toddler imaging~\citep{Barkovich2019-bm}}.}
\subsection{Importance of Longitudinal Changes}
Characterizing longitudinal changes is more present in other applications such as disease detection using normative modeling~\citep{Bethlehem2022-ni, Rutherford2022-sc, Chen2021-qs} in adult populations. For example, Alzheimer’s disease has been one of the most studied neurodegenerative diseases to characterize those morphological transformations happening through time~\citep{Gafuroglu2018-hb, Ouyang2021-jt, Ouyang2022-tr}. Usually, from the presence of atrophy measured by hippocampal volume or cortical thickness measures extracted from structural MRI images, normative measures can be distinguished from pathological cases~\citep{Jang2022-tp}. 

     \finalrev{The deformation field or} hidden transformation between two time points contains regions of contraction and expansion as well as potentially being a growth indicator. Hence, neurodevelopmental evolution trajectories could be extracted in the pediatric context. Indeed, changes are exponentially variable from 0 to 6 years of age, where a 6-year-old brain resembles at 95\% to an adult brain~\citep{Phan2018-ov}. Being dependent on the age interval, an analysis on deformable intra-subject registration performance can be of value to dissect developmental patterns. 

     \finalrev{The primary aim of this study is to conduct a detailed comparative analysis between SyN ANTs, a conventional registration algorithm and an unsupervised neural network. This involves evaluating the potential of a deep learning framework for intra-subject registration on pediatric longitudinal brain data when separating global and local transformations. Central to our objectives are: 1) examining the performance of unsupervised neural networks, with and without initialization registration tasks, and 2) analyzing how intra-subject age intervals and sex impact the performance metrics used to assess the DL-based registration. This comparative analysis will inspect the advantages and disadvantages of each method in terms of performance, training, age intervals, and inference times, thereby placing a significant focus on the distinctions between SyN ANTs and a Voxelmorph-like DL-based approach. Hence, one of the contributions of this study is to compare SyN ANTs~\citep{Avants2008} and DL-based methods in achieving a complete deformable registration task on pediatric longitudinal brain magnetic resonance (MR) images when using three pre-alignment approaches.} This work builds upon a previous short paper that was presented at the WBIR Workshop~\citep{Dimitrijevic2022-ns}. While the primary goal remains consistent, our approach differs in terms of the methodology, and we have introduced a comparative analysis with SyN ANTs, a conventional registration algorithm. The prior paper enabled a comparison between 1.5 mm and 2.0 mm isotropic images, revealing that the 1.5mm version yielded superior registration accuracy. Furthermore, we have expanded our investigation to incorporate age and sex-related analyses, capitalizing on the available longitudinal data to assess performance metrics with respect to age intervals in-between pairs. Additionally, this work includes an extended review of relevant literature.

% A methodological, model, or similar section often comes here.
\section{Methods}
\begin{figure}[!htb]
	\centering
	\includegraphics[width=1\linewidth]{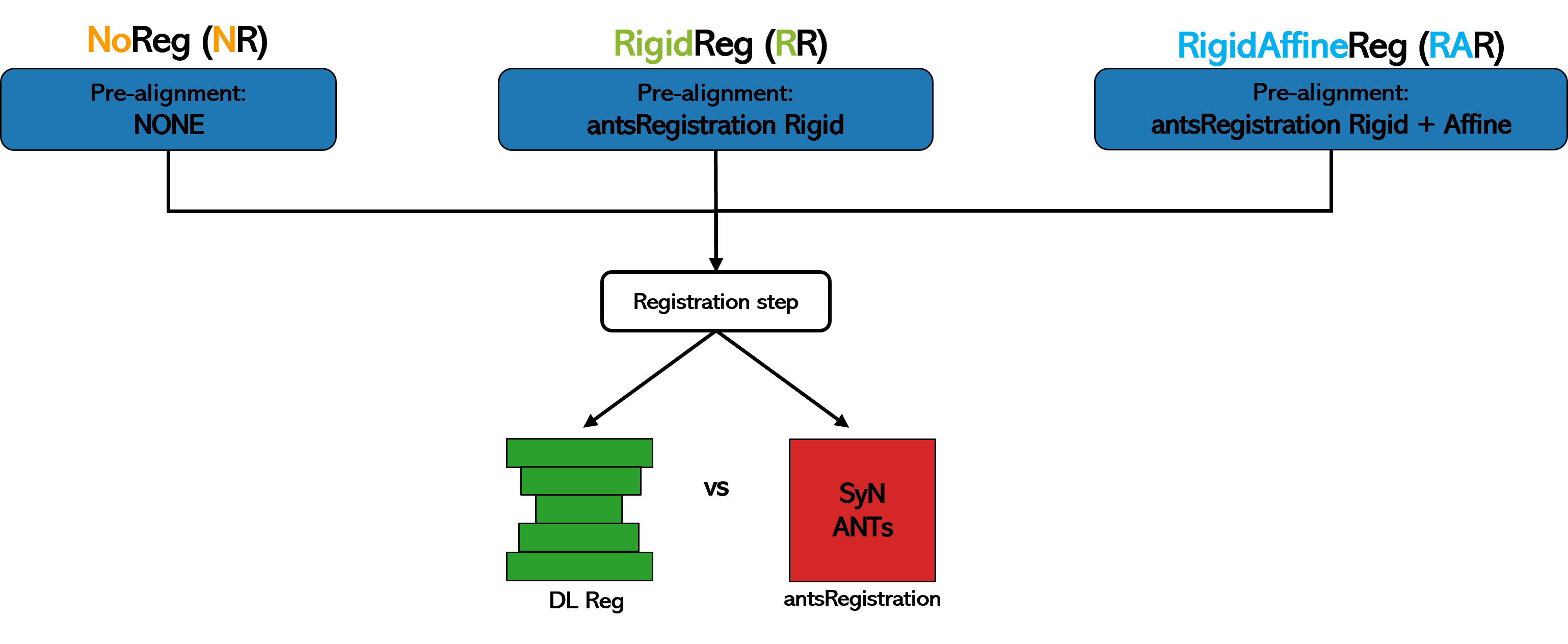}
	\caption{Illustration of the three \revision{initialization strategies, NoReg (NR), RigidReg (RR) as well as RigidAffineReg (RAR), (blue) used for comparing deep learning (green) and conventional SyN ANTs (red) registration approaches.}}
        \label{fig: method-scheme}
\end{figure}

\finalrev{As illustrated in Figure \ref{fig: method-scheme}, our investigation explores three initialization approaches: NoReg (NR), which involves no pre-alignment; RigidReg (RR), consisting of a rigid pre-alignment; and RigidAffineReg (RAR), incorporating both rigid and affine pre-alignments. These serve as inputs to either a U-Net for learning-based deformable registration (DL Reg) or SyN ANTs, serving as the conventional state-of-the-art comparison. SyN ANTs was selected due to its widespread use in the literature for various registration tasks in medical imaging~\citep{Tustison2021-am} and its benchmark performance in numerous competitions~\citep{Menze2015-ps,Murphy2011-gk}. On the other hand, the learning-based architecture and loss function was inspired by Voxelmorph~\citep{Balakrishnan2019-sr}, but specifically trained on pediatric data.} \revision{It is crucial to highlight that the U-Net architectures maintain consistent layer structures (as detailed in section \ref{chosen-architecture}), ensuring consistent comparison across all three initialization techniques.} The exact parameters used for rigid as well as rigid and affine initializations are \finalrev{available in section \ref{supp-material} in the supplementary material.} The reproducible pipeline is also available on the open-source Github repository at~\url{https://github.com/neuropoly/pediatric-DL-registration}.

% \subsection{Registration Approaches}
% \rev{\sout{The main contribution of this proposed method is the initialization steps which have been conducted prior to the neural network’s involvement with image pairs.}} \revision{\sout{Indeed, in this way, the impact of each type of transformation can be assessed by the decomposition of the full deformable transformation into simpler parts.}} \revision{As depicted in Figure \ref{fig: method-scheme}, three initialization approaches are illustrated: 

\subsection{Chosen Architecture} \label{chosen-architecture}
This subsection focuses on the unsupervised chosen DL framework to accomplish the full non-rigid registration task. Taking as input a moving (M) and fixed (F) image pair, the network computes a dense displacement field (DDF) which allows the creation of a \revision{moved image also called warped moving image or predicted fixed image}. This \revision{moved} image is obtained from aligning the moving image to the fixed image. All calculations are done with 3D \revision{volumes} as input. \revision{The selected CNN is a U-Net for which training optimizes a set of learnable parameters denoted as $\theta$, corresponding to the kernel weights of the network.} More specifically, the U-Net architecture which generates the deformation fields is a 3-layer encoder and decoder with 8, 16 and 32 channels each \revision{with skip connections. The specific parameters for the architecture are available in the config files on Github which are reproducible when using DeepReg.} A stochastic gradient descent method is used to find the optimal parameters of the network. In our case, an ADAM optimizer with a learning rate set to 1.0e-4 is used. Each training split was trained for 250 epochs and a batch size of 2 pairs of moving/fixed images. \finalrev{Figure \ref{fig: architecture} illustrates the loss function of the unsupervised network. This loss function comprises two terms: the first term, the local normalized cross-correlation (LNCC) similarity measure, selected for its robustness to local intensity variations. The second term represents L2-norm gradient regularization, with its weighting factor set to 1 based on an optimal value from the literature~\citep{Balakrishnan2019-sr}.} While exploring different L2-norm gradient regularization factors for realistic deformation fields would be insightful, this study's primary focus remains on assessing various initialization approaches. The training procedure and architecture initialization are facilitated by the version named develop 0.0.0 from DeepReg~\citep{RN38}, a DL-based registration framework. \finalrev{A 5-fold cross-validation scheme is used for train and test splits}. Finally, training is executed \rev{on a system equipped with Ubuntu 18.04.5 (amd64), featuring an octa-core Intel i7-9700KF CPU, 62.7 GB of RAM, and a GeForce RTX 2080 Ti GPU.}
\section{Data}
The chosen publicly available Calgary Preschool dataset~\citep{Reynolds2020-hq} was obtained using a General Electric 3T MR750w system and 32-channel head coil (GE, Waukesha, WI) at the Alberta Children’s Hospital in Calgary, Canada. This acquisition process received approval from the University of Calgary Conjoint Health Research Ethics Board. The dataset comprises T1-weighted MRI brain scans from 96 children aged 2 to 8 years. These scans were obtained using an FSPGR BRAVO sequence with the following parameters: TR = 8.23 ms, TE = 3.76 ms, TI = 540 ms, flip angle = 12 degrees, voxel size = 0.4492x0.4492x0.9 mm³, 210 slices, matrix size = 512x512, and a field of view = 23.0 cm. It includes multiple scans at different time points for 96 subjects, an essential element for single-modality intra-subject DL registration. \revision{Given the inherent challenges in obtaining pediatric datasets, the dataset size is particularly noteworthy, especially as it offers longitudinal samples~\citep{Lebel2018-gv}}. It also includes age, biological sex, handedness, and other parameters which can be further analyzed. From the 96 subjects, it was necessary to choose children with two or more time-point scans from the acquired data. Hence, 64 subjects respected those conditions which brings the used data to a total of 247 T1-weighted images allowing 434 combinations of moving/fixed registration pairs.

    Table \ref{tab: chosen-dataset} displays the dataset characteristics and relevant parameters. A graphical representation of the longitudinal age and sex parameters is also available in Figure \ref{fig: data-info}. Additionally, the remaining 64 subjects were inspected for image quality. \revision{Each image in the dataset was verified to possess the stated matrix dimension of 512x512x210.} \finalrev{For the data preprocessing details, refer to section \ref{supp-material} in the supplementary material. A full pipeline including steps from data preprocessing to training is also available \rev{in} Figure \ref{fig: full-pipeline} in the supplementary material.}

    \begin{table}[!htb]
    \centering
    \caption{Characteristics of the chosen subset from the Calgary Preschool dataset. SD is for standard deviation.}
    \vspace*{0.25cm}
\begin{tabular}{lc}
\toprule
            &   \textbf{Calgary Preschool subset} \\ \midrule
\textbf{No. of subjects}  &  64\\\addlinespace
\textbf{No. of images} &   247 \\\addlinespace
\textbf{No. of scans/subject} & \\\addlinespace
\hspace{1cm} \textbf{Average (SD)} &  3.86 (1.59) \\\addlinespace
\hspace{1cm} \textbf{Min} &  2 \\\addlinespace
\hspace{1cm} \textbf{Max} &  10 \\\addlinespace
\textbf{Total no. of possible combination pairs} &   434 \\\addlinespace
\textbf{Age of all time point scans (years)} & \\\addlinespace
\hspace{1cm} \textbf{Average (SD)} &  4.49 (1.00) \\\addlinespace
\hspace{1cm} \textbf{Min} & 1.97 \\\addlinespace
\hspace{1cm} \textbf{Max} &  6.90 \\\addlinespace
\textbf{Sex no. of (\%)} &    \\\addlinespace
\hspace{1cm} \textbf{Female} &  107 ($\sim$43) \\\addlinespace
\hspace{1cm} \textbf{Male} &  140 ($\sim$57) \\\addlinespace
\textbf{Handedness no. of (\%)} &    \\\addlinespace
\hspace{1cm} \textbf{Right} &  214 ($\sim$87) \\\addlinespace
\hspace{1cm} \textbf{Left} &  25 ($\sim$10) \\\addlinespace
\hspace{1cm} \textbf{Both} &  8 ($\sim$3) \\\addlinespace
\bottomrule
\end{tabular}
\label{tab: chosen-dataset}
\end{table}

\section{Experiments}
\subsection{Segmentations Retrieval}
\revision{\rev{For} segmentations, $SynthSeg^+$~\citep{Billot2023-po}, also referred as SynthSeg in the following analyses, was used to obtain 32 labels representing various brain structures, resulting in a total of 18 regions for segmentation. \finalrev{Subsequent analyses focus on both the 18 regions individually and globally on three primary tissues: white matter (WM), gray matter (GM), and cerebrospinal fluid (CSF). For global tissues,} WM encompasses cerebral white matter, brain stem, cerebellum white matter, pallidum, and ventral DC; GM includes cerebral cortex, crebellum cortex, accumbens area, caudate, thalamus, putamen, hippocampus, and amygdala; and CSF comprises the lateral ventricle, inferior lateral ventricle, 4th ventricle, 3rd ventricle and CSF regions. \rev{For QC, initially, three pairs out of the 64 available pairs were excluded using SynthSeg's QC scores, as one subject within each pair had values lower than 0.6~\citep{Billot2023-ch}. Subsequently, a visual QC check was performed, and no images were excluded.}}
\subsection{Validation Process}
\finalrev{Validating registered images is challenging due to the absence of common and normalized decision metrics~\citep{Christensen2006-ja}. In this work, we inspect two types of Dice scores and the influence of age interval per pair and sex-specific differences on performances, pro-rated training and inference times, percentage of negative Jacobian determinant (JD) and sum of absolute log JD values. A detailed overview is present for each metric below. Additionally, the equations used for Dice scores and JD related metrics are detailed in section \ref{metrics} in the supplementary material.}

\begin{figure}[!htb]
	\centering
	\includegraphics[width=\linewidth]{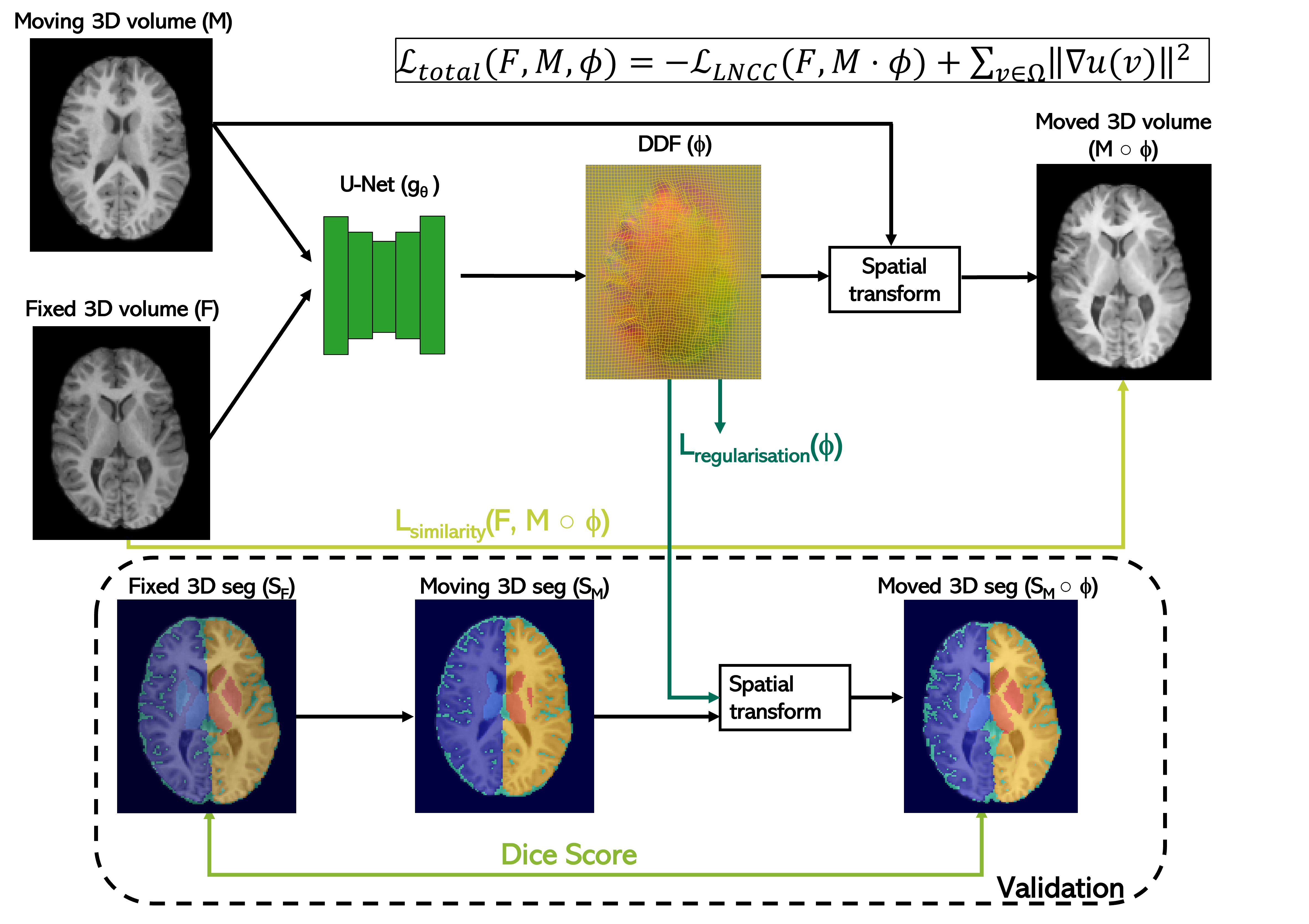}
	\caption{Schema of the training procedure to obtain \finalrev{a} deformation field ($\phi$) with given moving (M) and fixed (F) 3D pair of images. The validation technique using WM, GM and CSF segmentations (WM depicted in this figure) to calculate Dice scores is also shown in the \revision{black} \revision{dashed} region as well as the loss function in the upper right corner \finalrev{where $v$ indicates voxels for the L2 norm of the displacement gradient, $\nabla u$, which encourages a smooth deformation field. $\phi$ is calculated by adding the identity transform to the displacement field ($\phi = Id + u$).} Image inspired by~\citep{Balakrishnan2019-sr}.}
        \label{fig: architecture}
	\end{figure}

    \finalrev{\textbf{Unweighted and Weighted Dice Scores:}} As depicted in Figure \ref{fig: architecture}, the validation process in terms of Dice scores is represented in black \rev{dashed} lines. Indeed, moving segmentations are warped using the network’s output DDFs and then compared to the fixed segmentations with the Dice score as the performance metric. Maximal region overlap is indicated by a Dice score of 1 and no overlap by 0. \rev{It's crucial to highlight that when averaging Dice scores across regions of different sizes, larger regions like CSF, cerebral white matter, and cerebral cortex may dominate the results. Indeed, overlap scores of small and localized anatomical regions are to be prioritized as reliable indicators that can differentiate between plausible and inaccurate registrations~\citep{Rohlfing2012-zj}. \finalrev{To reduce the influence of larger regions, a weighted Dice score is calculated with the weight consisting of the inverse number of voxels per region. This weighted Dice score is used to assess performances for each one of the 18 available regions. SyN ANTs and DL-based approaches were also compared in terms of the commonly used Dice score in the literature, referred as unweighted Dice score, on each of the 18 regions and on WM, GM and CSF \rev{by averaging sub-regions within these tissues from the total 18 regions available from $SynthSeg^+$~\citep{Billot2023-po}}. \finalrev{Unweighted Dice scores for each of the 18 regions and for the three global tissues are plotted against age intervals in years per pair to measure the influence of age. Additionally, the same age-related analyses are performed separately for each sex.} 
    
    \textbf{Pro-rated Training and Inference Times:} In our evaluation, we employ the concept of pro-rated training time as a key metric for assessing the computational efficiency of model training relative to dataset size. Pro-rated training time is calculated by dividing the total training time by the number of pairs present in the dataset. Regarding inference times, they denote the duration required to generate a warp field, either through prediction using the trained U-Net model or via running the SyN ANTs algorithm for each pair. They were performed on each pair five times and averaged across the five folds.
    
    \textbf{Negative JD and Sum of Absolute Log JD values:} Calculated deformation fields \rev{from both DL-based and SyN ANTs registration methods} were evaluated for foldings using the percentage of negative JD values \rev{and the average sum of absolute log JD for all initialization approaches.} \finalrev{These metrics assess invertibility and local topology preservation of deformation fields, which are crucial for adequate registration quality. To verify if those properties are ensured, the percentage of negative JD and the sum of absolute log JD values are computed on each DDF generated per registration pair. A negative JD indicates non-invertibility properties due to the presence of unwanted local folding. \finalrev{As for log normalized JD values, they show volumetric changes post-registration: negative values indicate local volume contraction, while positive values indicate local volume expansion.} For the average sum of absolute log JD over all pairs, it is calculated within local sub-regions extracted from SynthSeg and encompassing global WM, GM, and CSF regions, excluding background regions. Finally, the distribution of the sum of absolute log JD values within the whole brain region is inspected for both DL Reg and SyN ANTs to compare their regularization strengths. These metrics offer insights into the deformation characteristics within specific brain regions and potential distortions, helping assess deformation smoothness or regularization strengths between methods.}}}

\section{Results}
\begin{figure}[!htb]
	\centering
	\includegraphics[width=1\linewidth]{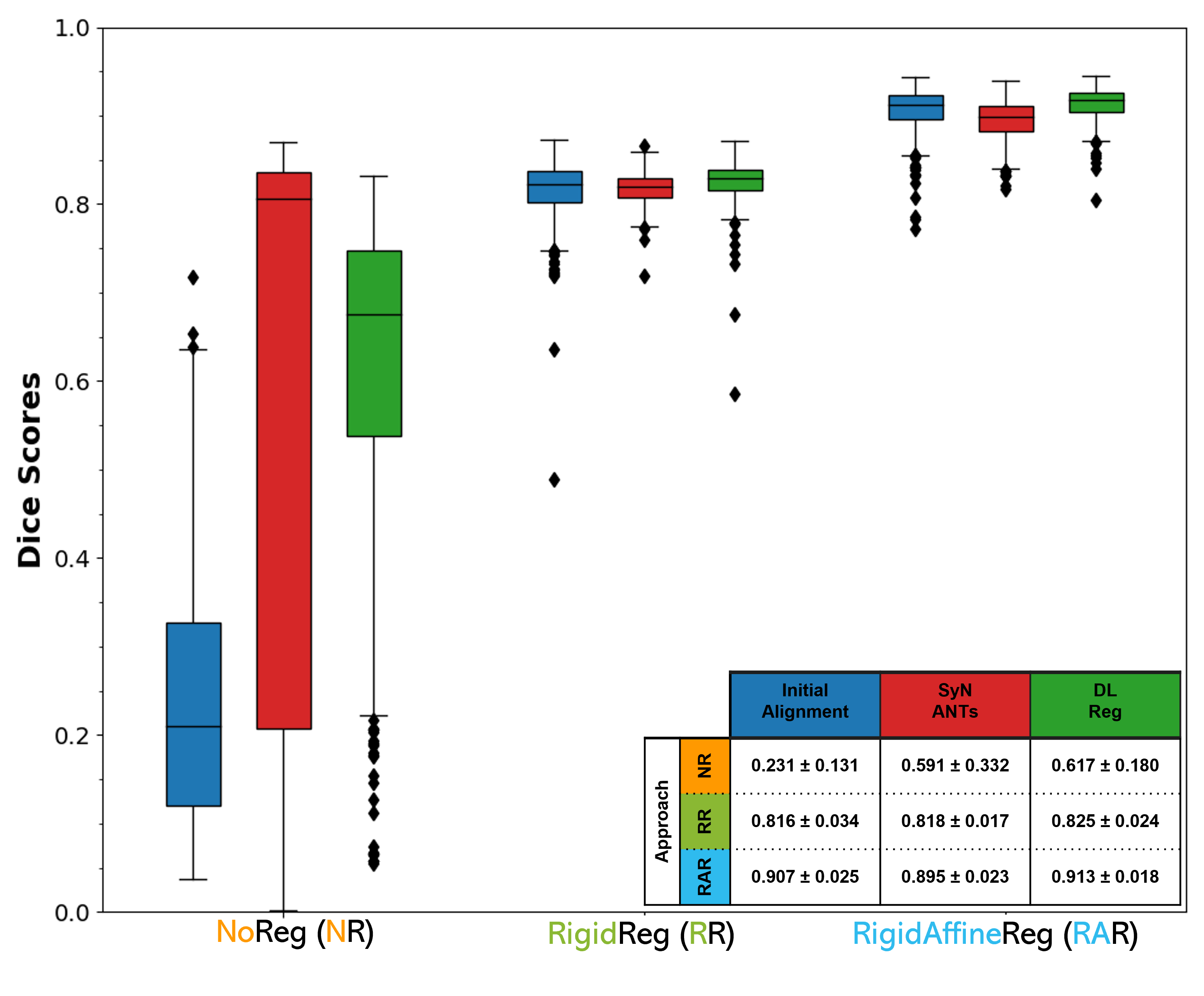}
	\caption{Dice score results on the test sets represented as boxplots for each \revision{initialization} approach (NoReg, RigidReg and RigidAffineReg) for DL-based methods compared to the initial \rev{Dice} scores pre-conducting the registration steps. Each method is also compared to the SyN ANTs registration. The \rev{Dice} scores are averaged over \revision{all 18 segmented regions. The table in the lower right corner shows the \rev{mean±SD} Dice scores for all scenarios.}}
        \label{fig: dice-results}
	\end{figure}

Figure \ref{fig: dice-results} illustrates Dice scores averaged across \revision{all 18} segmented regions for the three initialization methods: NR, RR, and RAR \revision{for DL-based registration} (in green), contrasted with the results obtained using SyN ANTs (in red). Initial Dice scores prior to \rev{both SyN ANTs and} U-Net processing are shown in blue. \rev{Dice scores weighted based on the inverse number of voxels per region and normalized by the sum of weights for all 18 regions from SynthSeg are available on supplementary Figure \ref{fig: dice-results-weighted} which exhibit similar trends as in Figure \ref{fig: dice-results}.} For comprehensive Dice score breakdowns across the segmented regions and comparisons to SyN ANTs, refer to Table \ref{fig: table-final} and Figure \ref{fig: dice-per-region} \revision{for results across WM, GM and CSF} \rev{obtained by averaging
SynthSeg sub-regions within these tissues from the total 18 regions available.} \rev{Figure \ref{fig: dice-per-all-regions} in the supplementary material also shows Dice scores for all 18 segmented regions separately averaged over all subjects for all three initialization scenarios.} Additionally, Table \ref{fig: table-final} presents a comparison \revision{of pro-rated training time and inference} durations between DL-based and ANTs registration pipelines. Notably, \finalrev{for all three initialization approaches, the DL-based method demonstrates} markedly faster \revision{inference} times per pair when the model is trained, between 22 to 74 times faster, compared to the required \rev{SyN} ANTs registration time. \rev{However, when comparing pro-rated training times to the same SyN ANTs registration times per pair, SyN ANTs is around 1.4 times faster for both RR and RAR initializations.}

    Furthermore, due to the non-parametric nature of the data, two-sided Wilcoxon tests have been conducted \revision{for all three initialization approaches between DL Reg and SyN ANTs}. \finalrev{For NoReg, the median results from DL-based approaches are not statistically different from those of SyN ANTs across all three segmented regions (p-values $>$ 0.18, p $>$ 0.05). In contrast, for RigidReg and RigidAffineReg, the results from DL-based approaches show a statistically significant difference from those of the corresponding ANTs pipelines in all three regions (p-values $<$ 1.27e-17, p $<$ 0.05).} As shown in \finalrev{Table} \ref{fig: table-final}, for both RR and RAR initializations, DL Reg outperforms SyN ANTs for all three segmented regions. \finalrev{However, SyN ANTs has decreased performances compared to the initial alignment}. As for NR, only the average Dice score calculated on CSF tissues \rev{for DL Reg} are equal in comparison to SyN ANTs however with a smaller standard deviation. \finalrev{When examining each of the 18 segmented regions individually, p-values and statistical significance (p$<$0.05) were determined from Wilcoxon tests, which compare SyN ANTs Dice scores to initial Dice scores before any initialization (Figure \ref{fig: ANTs_init_stats}), DL Reg Dice scores to initial Dice scores before any initialization (Figure \ref{fig: DL_init_stats}) and DL Reg Dice scores to SyN ANTs Dice scores (Figure \ref{fig: DL_ANTs_stats}), all of which are presented in the supplementary material. Learning-based results show a statistically significant difference across all 18 regions when compared to the initial alignment for NR. However, there are some exceptions for the two other initialization approaches: for RR initializations in regions such as the caudate, 3rd ventricle, and amygdala, and for RAR initializations in the putamen, pallidum, 3rd ventricle, hippocampus, amygdala, and accumbens area. Also, the percentages of instances where SyN ANTs outperforms DL Reg for NR (64.1\%), RR (28.6\%), and RAR (10.8\%) initializations were calculated across all 18 regions, considering a total of 431 pairs.} 
    
    \begin{table}[!htb]
  \caption{Dice scores and \revision{average sum of absolute log JD values} per segmented regions, white matter (WM), gray matter (GM) and cerebrospinal fluid (CSF) \revision{by averaging SynthSeg sub-regions within
these tissues from the total 18 regions available} for all three proposed approaches and their comparison to SyN ANTs \revision{as well as the initial alignment.} \revision{This average is calculated by dividing by the number of voxels per region.} \revision{Pro-rated training (divided by the number of pairs) and inference} registration time per pair are shown. Values are presented as mean ± SD.}
  \includegraphics[width=\linewidth]{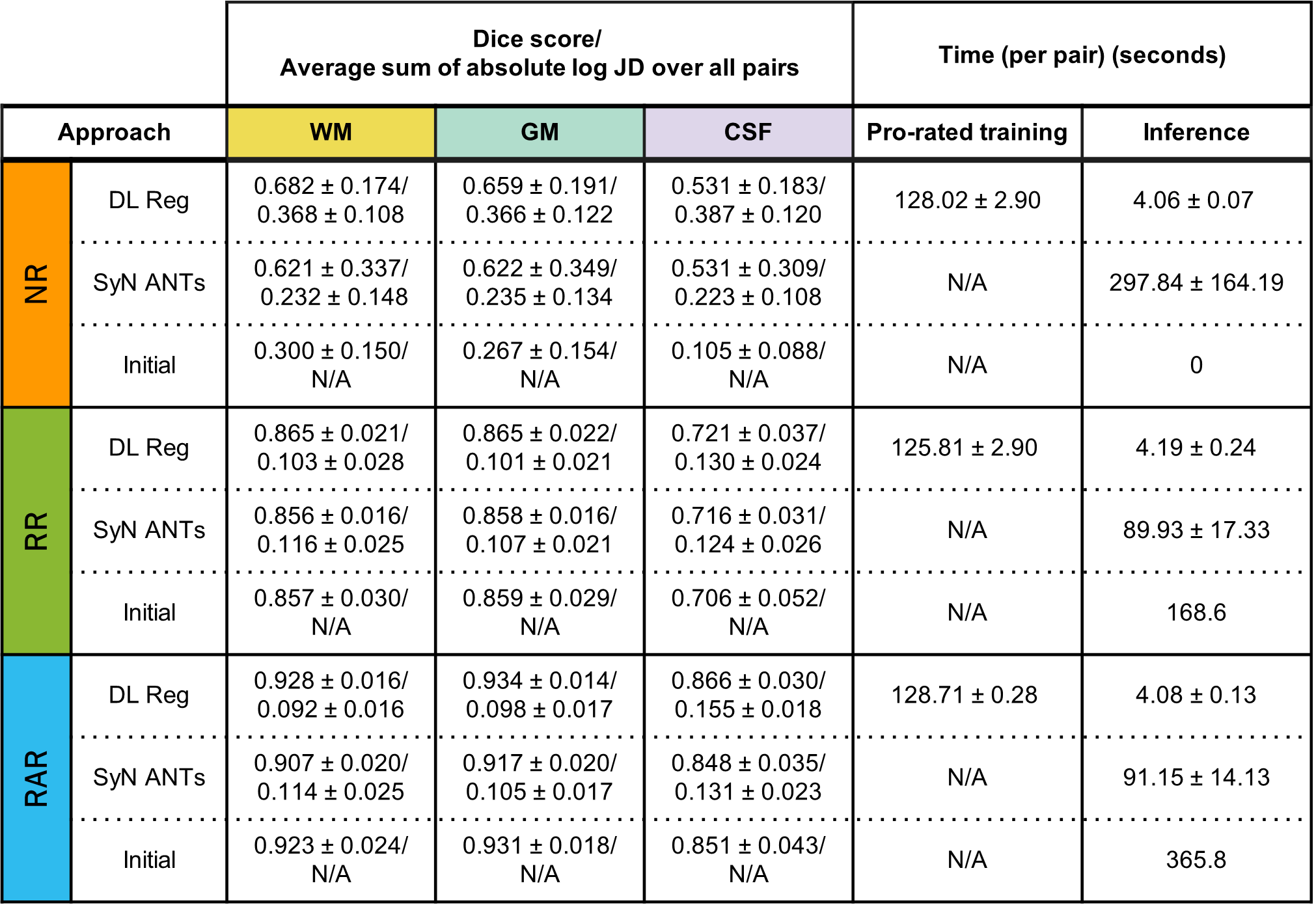}
  \label{fig: table-final}
\end{table}

\begin{figure}[!htb]
	\centering
	\includegraphics[width=1\linewidth]{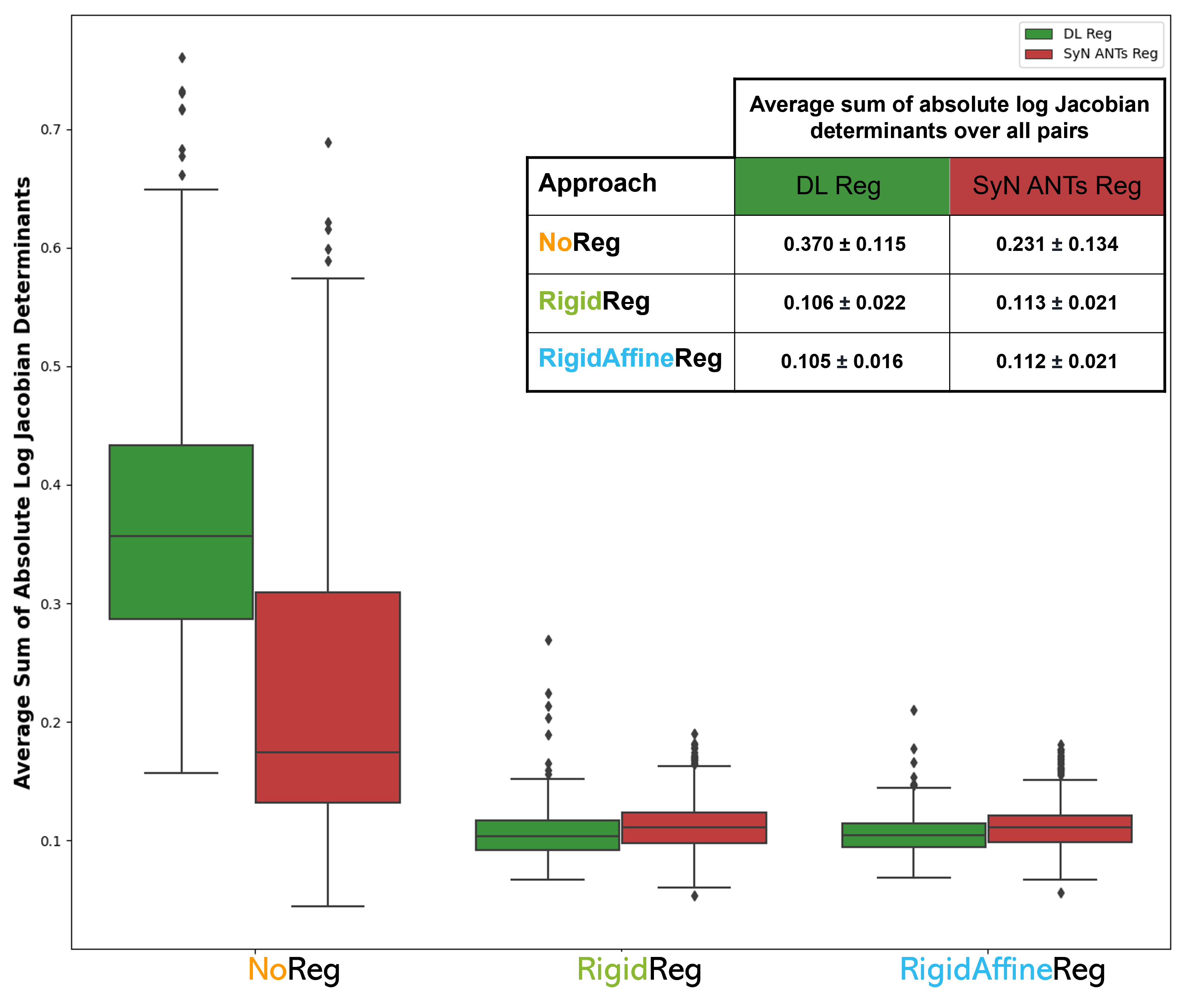}
	\caption{\revision{Average sum of absolute Log Jacobian determinants for all three initialization approaches with DL-based in green and SyN ANTs in red. This average is calculated by dividing the sum by the total number of voxels in the brain region per volume. In the upper right corner, a table of the average over all pairs is presented.}}
        \label{fig: sum-jacobians-global}
	\end{figure}

    Also, Table \ref{fig: table-final} highlights the \revision{average sum of absolute log JD values for all three segmentation regions for both DL-based and SyN ANTs approaches. \revision{\finalrev{Also, it is to be noted that} the percentage of negative JD values is 0\% for all initialization approaches.} As visible on Table \ref{fig: table-final}, NR contains the larger sum of absolute log JD values for all segmented regions\finalrev{,} but remains lower than 0.387. DL Reg seems to lead to more unrealistic deformations compared to SyN ANTs when no prior intialization is done. This is the opposite when looking at RR and RAR initializations, where DL Reg has on average smaller values than SyN ANTs except in the CSF region. It suggests that the deformation fields are relatively smoother for DL Reg whilst still achieving higher Dice scores. As expected, RR and RAR seem to be more robust to local foldings as their average sum of absolute log JD over all pairs, no matter the segmented region, are always below 0.155 for both DL Reg and SyN ANTs.} \revision{For further regularization analysis, boxplots in \rev{Figure \ref{fig: sum-jacobians-global}} depict the distribution of the sum of absolute log \finalrev{JD} values within the whole brain region. This graphical representation highlights that, with the exception of NR, both DL-based registration and SyN ANTs exhibit comparable average absolute sums of log \finalrev{JD} for all initialization methods. In the case of NR, DL Reg shows larger values, just as the trends observed in segmented tissues. Lastly, refer to supplementary Figure \ref{fig: example-fields} for an illustration of warped segmentations and their corresponding warping fields for each initialization method and DL-based registration, allowing for a visual examination of their smoothness.}

\begin{figure}[!htb]
	\centering
	\includegraphics[width=1\linewidth]{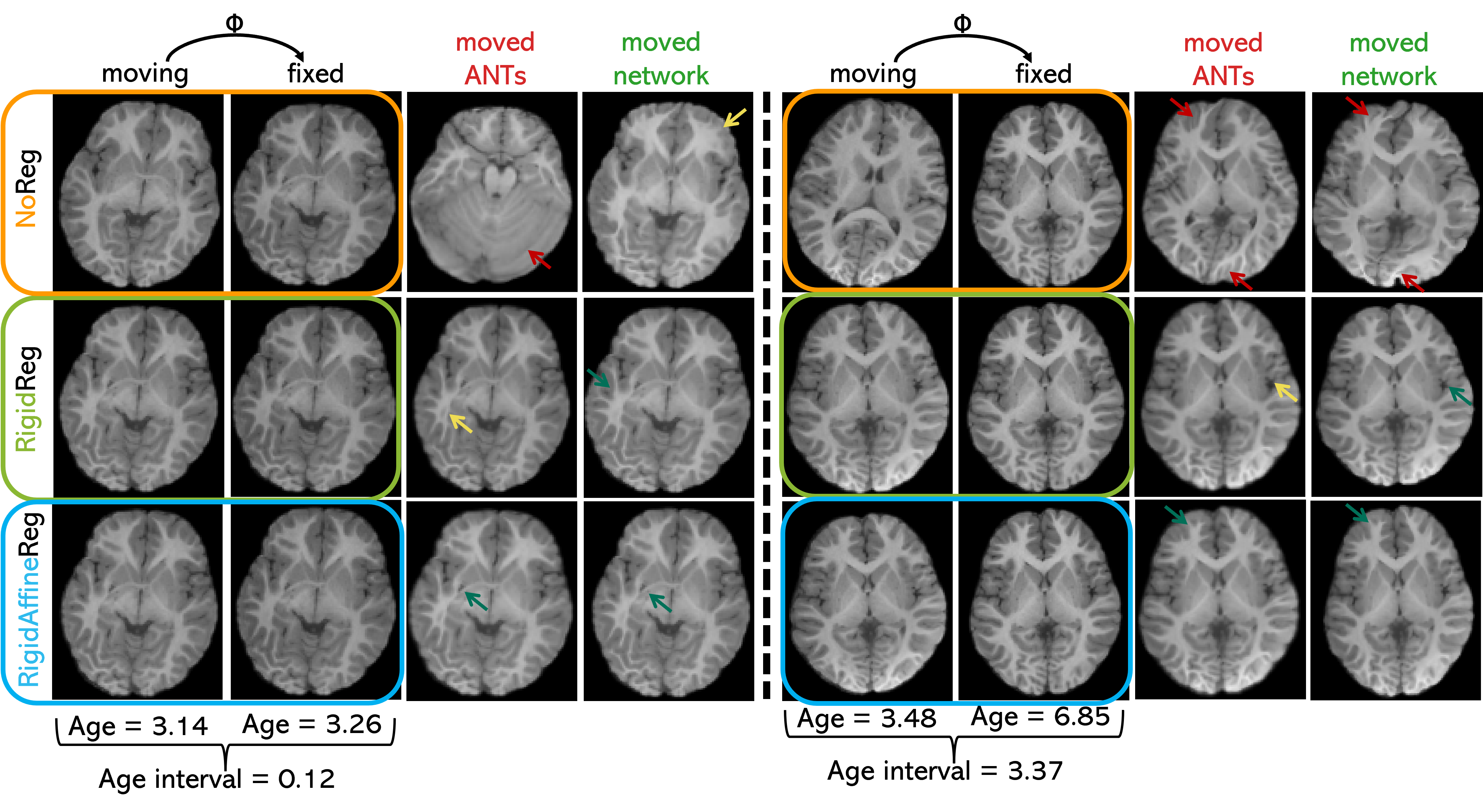}
	\caption{Visual representation of the results obtained with all three DL-based approaches (\revision{moved} network) compared to SyN ANTs (\revision{moved} ANTs) results with an age interval between moving and fixed images of 0.116 years on the left and 3.37 years on the right. Red arrows highlight instances of misalignment, yellow arrows indicate blurriness or minor deviations from the fixed image, and green arrows denote successfully aligned areas.}
        \label{fig: network-results}
	\end{figure}

        Figure \ref{fig: network-results} displays an example of registration results on two subjects for all three proposed approaches as well as SyN ANTs counterparts. \revision{Moved} images are obtained by warping the moving image with the obtained DDFs \revision{for both DL-based and SyN ANTs approaches using antsApplyTransforms function from ANTs}. Red arrows in the images indicate regions where the warping did not go as expected. On the other hand, green arrows indicate relatively good reproductions of the desired fixed image when inspecting the \revision{moved} image. Yellow arrows identify areas in proximity to the structures from the fixed image. However, these areas may display blurriness or slight deviations from the expected alignment, indicating regions that are not perfectly matched. For both age intervals, \rev{in the} RR and RAR \rev{scenarios, DL Reg seems} to attain similar results as SyN ANTs visually. 

\begin{figure}[!htb]
	\centering
	\includegraphics[width=1\linewidth]{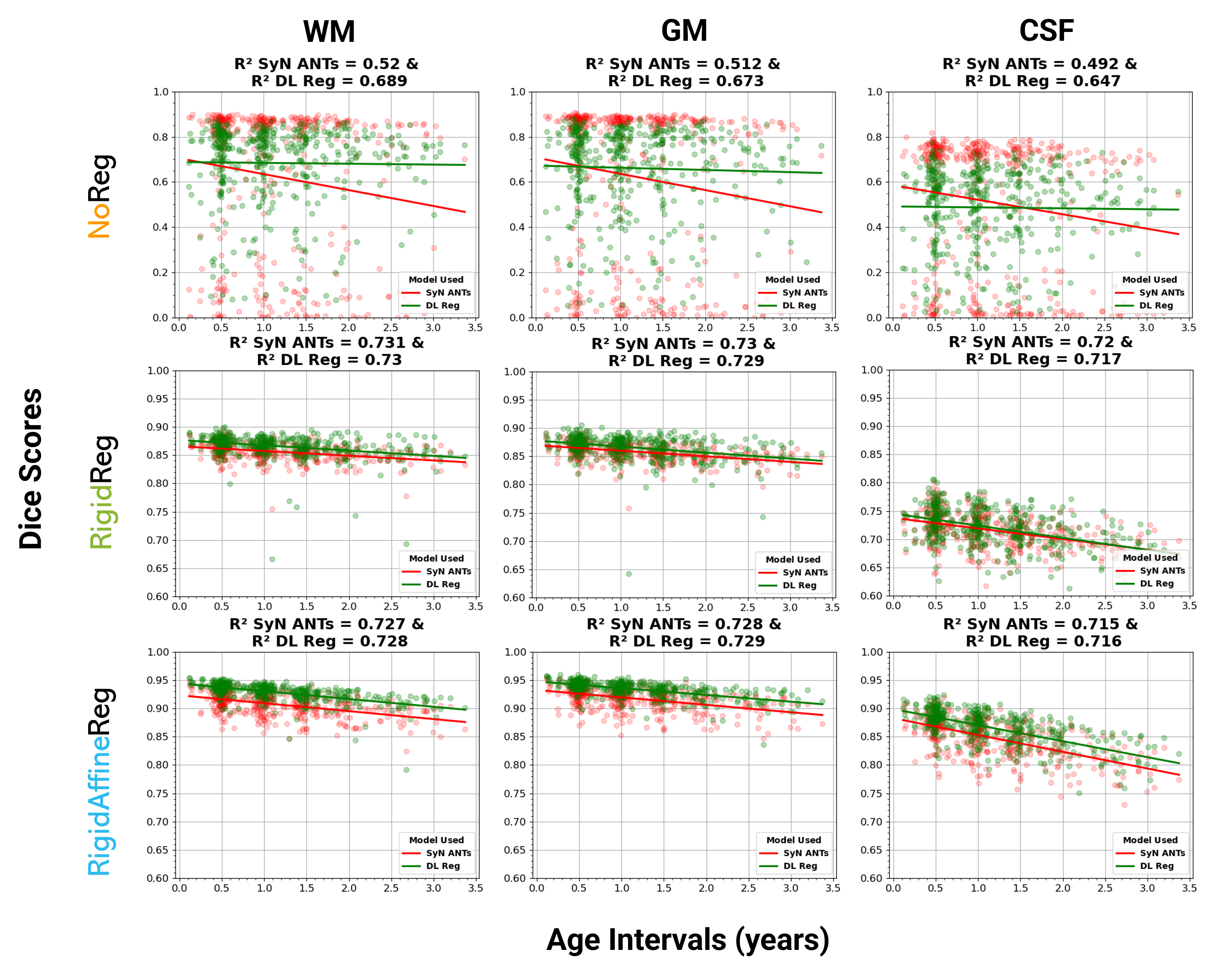}
	\caption{Dice scores against age intervals for all initialization methods compared to SyN ANTs. The three rows correspond to NoReg, RigidReg, and RigidAffineReg, and the three columns represent WM, GM, and CSF segmentations. SyN ANTs Dice scores are shown in red, while the results of DL-based approaches are in green, with corresponding trendlines in the same color. \rev{Dice scores in global regions are calculated by averaging SynthSeg sub-regions within these tissues from the total 18 regions available}. Figure titles include coefficients of determination (R-squared) for reference. Note: the y-axis scale for NoReg differs due to a distinct range of Dice scores.}
        \label{fig: dice-vs-age}
	\end{figure}

To better consider the influence of age intervals on Dice score outcomes, refer to Figure \ref{fig: dice-vs-age}, which depicts Dice scores for individual segmented regions against age intervals measured in years per pair. The green data points and corresponding trendlines illustrate outcomes from the proposed initialization followed by DL-based registration. These are contrasted with SyN ANTs Dice scores, represented by red data points and trendlines \revision{for all three initialization approaches considered: NR, RR and RAR}. The trendlines, visualized through linear regressions, were obtained using ordinary least squares estimation, with the R-squared value reflecting squared linear correlation. In the learning-based context, the average coefficient of determination stands at \revision{0.707}. This signifies that, on average, more than \revision{70.7\%} of the explained Dice score variance is attributed to age interval fluctuations, irrespective of the initialization method employed. Comparatively, the state-of-the-art SyN ANTs pipeline exhibits a slightly lower percentage, with an average R-squared value of \revision{0.653}. This discrepancy is primarily due to diminished coefficients when no initialization is applied. In terms of performance, the green trendlines for RR and RAR approaches consistently surpass the red counterparts associated with the conventional ANTs pipeline, across all three segmented regions, except for \revision{the CSF and GM in the NR approach}. For smaller age intervals, SyN ANTs demonstrates enhanced performance in this \revision{particular} scenario\revision{, specifically for an age interval less than 0.6 for GM and less than 1.6 for CSF. A Plotly HTML displaying trendlines of Dice scores versus age intervals for \rev{each one of the} 18 segmented regions can be accessed on the GitHub repository.} 

    The same analyses were performed by separating sex information and seeing if it has an impact per initialization approach. These are available in supplementary figures \ref{fig: noreg-sex} through \ref{fig: rigidaffinereg-sex}. 

\section{Discussion}
In this discussion, we examine the registration results \finalrev{obtained using the conventional SyN ANTs alignment method and the U-Net learning-based method.} Initially, we analyze Dice scores across \revision{all 18 segmented regions\rev{, and afterward, we focus on sub-regions within these 18 regions that are included in WM, GM and CSF global tissues}. Then, the average sum of absolute JD values are inspected for the whole brain, but also per segmented tissues.} Subsequently, we highlight the disparities in \rev{both pro-rated training and} inference times required for registering intra-subject pairs. Moving forward, we inspect age-related analyses, considering age intervals per pair and any sex-specific differences. We finish by touching upon the generalization capacities of each registration approach \revision{and some recommendations for applying intra-subject registration in the pediatric context}.

     \finalrev{\textbf{Dice Scores Across All 18 Regions:}} When applying a pre-alignment strategy before undergoing an unsupervised U-Net, both RR and RAR DL-based approaches exhibit similar performances to the conventional ANTs algorithm in terms of Dice scores. However, when the network is trained, they are faster in terms of registration times compared to the conventional approach. These two DL-based approaches can register two intra-subject images with great quality proven by the high average Dice score results obtained (\revision{0.825±0.024 and 0.913±0.018} for RR and RAR respectively) as seen \rev{in} Figure \ref{fig: dice-results} whilst operating 22 times faster than the SyN ANTs algorithm. \rev{Nevertheless, when comparing pro-rated training times to the corresponding SyN ANTs registration times per pair, SyN ANTs demonstrates approximately 1.4 times greater speed for both RR and RAR initializations.} \revision{For the NR approach, SyN ANTs achieved an average Dice score of 0.591±0.332, and DL Reg achieved 0.617±0.180. Both are higher than the initial Dice score of 0.231±0.131. \rev{Interestingly, although SyN ANTs exhibited a lower mean Dice score, its higher median Dice score (0.806) compared to DL Reg (0.676) suggests a more robust performance with occasional complete registration failures, highlighting the trade-off between consistency (higher mean) and the potential for superior performance (higher median). It's also crucial to note that for NR, there is no statistical difference between SyN ANTs and DL Reg median Dice scores. This observation may explain why, in Figure \ref{fig: dice-results-weighted}, the only initialization for which the trend differs from Figure \ref{fig: dice-results} is NR. In this case, when more weight is given to smaller regions based on their number of voxels, SyN ANTs has a mean Dice score of 0.532±0.331, while DL Reg has a mean Dice score of 0.488±0.205, even though SyN ANTs maintains a higher median weighted Dice Score (0.73). For RR, SyN ANTs has the same mean Dice score of 0.735 as after initial alignment with a smaller standard deviation. Whilst for the unweighted Dice score,} SyN ANTs achieved a Dice score of 0.818±0.017, and DL Reg achieved 0.825±0.024 \rev{for the RR approach}. Again, both are higher than the initial Dice score of 0.816±0.034. Finally, for the RAR approach, SyN ANTs achieved a Dice score of 0.895±0.023, and DL Reg achieved 0.913±0.018. This time, only the DL Reg Dice score is higher than the initial Dice score of 0.907±0.025, not SyN ANTs. It is noteworthy that while SyN ANTs has lower Dice scores for RR and RAR initializations, it demonstrates fewer outliers compared to DL Reg and the initial alignment. Examining more localized regions as illustrated in Figure \ref{fig: dice-per-all-regions}, DL Reg exhibits slightly superior performance compared to SyN ANTs across all 18 regions. Notably, CSF, the 3rd ventricle, and the 4th ventricle show the lowest Dice Scores, while the hippocampus, caudate, and putamen exhibit the highest scores. Frequently, in the context of RR and RAR initializations, SyN ANTs fails to achieve higher Dice scores than those obtained during the initial alignment. Particularly in the case of RAR initialization, where images share substantial similarity in intra-subject registration, SyN ANTs struggles to detect local changes associated with neurodevelopment. This suggests that SyN ANTs may benefit from more nuanced fine-tuning tailored to specific age ranges or individual pairs of images, rather than employing a global approach across a target population. \rev{Indeed, to fine-tune ANTs more appropriately for specific age ranges, the primary approach involves using distinct registration parameters tailored to the age group and individual in question. This, however, entails a lengthy and manual process. In contrast, DL-based registration streamlines this by enabling the network to learn and adapt through the training process, eliminating the need for manual adjustments specific to age ranges, although it is trained only for a specific population.} \finalrev{Considering the instances where SyN ANTs outperforms DL Reg with NR (64.1\%), RR (28.6\%), and RAR (10.8\%) initializations across all 18 regions and 431 pairs, a hybrid approach could be beneficial. Initially, employing the DL-based approach for all pairs, followed by using SyN ANTs for cases where the registration is unsuccessful, could mitigate processing time while ensuring satisfactory registration results.} It is also crucial to note that while all segmentations generated using the DL-based SynthSeg segmentation method underwent quality control evaluation, they still do not constitute ground truth labels. Nevertheless, the method's robustness is evident as it was trained and validated on diverse clinical datasets~\citep{Billot2023-ch}.} 
     
    \finalrev{\textbf{Dice Scores in Global Tissues:}} \revision{When examining each segmented tissue separately within the smaller set of 18 regions derived from SynthSeg, where WM and CSF each constitute 27.8\% of the total regions, and GM makes up the remaining 44.4\%, DL Reg using the RAR initialization outperforms RR for WM (0.928±0.016 vs. 0.865±0.021), GM (0.934±0.014 vs. 0.865±0.022) and CSF (0.721±0.037 vs. 0.866±0.030) segmentations. However, CSF Dice scores consistently tend to be lower compared to WM and GM for both DL Reg and SyN ANTs across all three initialization approaches. Our interpretation is that CSF is more susceptible to the effects of minor misregistration, given its predominantly thin structure around the brain, encompassing the CSF and ventricles. In contrast, WM and GM comprise larger, internal structures, such as the brain stem and cerebral white matter \rev{within} WM, and the cerebellum cortex \rev{within} GM.} These superior results from RAR are to be expected as all global transformations (rotation, translation, shear, and scaling) are accounted for by the rigid and affine registration pre-alignment steps used for RAR. Therefore, as visualized in Figure \ref{fig: network-results}, the RAR method can uncover local modifications on the \revision{moved} image after passing through the network. Contrary to RAR, NR has difficulties doing both local and global transformations as no \revision{initialization} was applied. Indeed, as presented \rev{in} Figure \ref{fig: network-results}, parts of the brain microstructures are an unrealistic warping combining both elements from the moving and fixed images for the example pair with an age interval of 3.37 years. The local transformations are not reproduced as well as when using the two other initialization approaches (RR and RAR). Indeed, as visible on both examples in Figure \ref{fig: network-results}, using only a SyN transformation does also not reproduce smaller microstructures correctly, sometimes completely warping incorrectly (for the lower age interval), as no global transformation is done beforehand. 

     \finalrev{The lower Dice scores for both RR and RAR initialization for each of the 18 regions and looking at the three global tissues may be attributed to the non-optimization of SyN ANTs in a population-based manner, unlike the network. Indeed, the network-based registration relies on learning optimal parameters through training from the provided input population. In contrast, SyN ANTs necessitates individual tweaking for each pair, demanding more human effort. Therefore, the key takeaway is that both SyN ANTs and DL Reg require parameter optimization, but the chosen method varies based on whether registration needs to be performed at a larger scale (encompassing all local regions and multiple subjects) or more locally for a specific small structure. Network-based registration proves fast at inference time, adapting quickly to a given input population, but requires training, potentially adding processing time while being more accurate for the studied population. On the other hand, SyN ANTs requires manual fine-tuning, and clear guidelines are lacking, particularly for younger cohorts, regarding whether to choose age-specific parameters or normalize across ages.}
    
    \finalrev{\textbf{Negative JD and Average Sum of Absolute JD Values:}} As for the percentage of negative JD values, evaluating if the deformation fields are realistic, the regularization factor of one seems to be sufficient in keeping this percentage \revision{to zero} for all three approaches. \rev{The findings regarding the average sum of absolute log JD values indicate comparability in regularization strengths between DL Reg and SyN ANTs methods, both within the brain region (Figure \ref{fig: sum-jacobians-global}) and for each segmented region (Table \ref{fig: table-final}).} Part of the success in DL-based approaches can be attributed to the fact that unsupervised learning is used instead of supervised learning. Indeed, supervised learning, particularly the attempt to predict ANTs-generated deformation fields, might have imposed restrictions on DL-based algorithms by confining their evaluation solely to a comparison with pre-existing ANTs-generated deformation fields.
    
        \finalrev{\textbf{Pro-rated Training and Inference Times:}} For the required \finalrev{inference} times per pair presented in Table \ref{fig: table-final}, the DL-based methods notably outpace the conventional ANTs pipelines, achieving a speedup of 22 to 74 times once the models are trained. \revision{However, the slightly improved Dice scores for RAR initialization compared to RR, observed in both DL Reg and SyN ANTs, come at the cost of doubling the initialization registration times.} \rev{Concerning the pro-rated training time for DL Reg, calculated by dividing the total training time by the number of training pairs, it averages around 127.51±2.37 across all initialization approaches. In contrast, SyN ANTs doesn't undergo training but necessitates parameter adjustments when a predefined set fails to achieve optimal performance. This process lacks explicit guidelines, usually initiating with default parameters or values recommended by the literature for the specific population dataset, and subsequently adjusting parameters per pair through quality control procedures.}
        
            \finalrev{\textbf{Age-related Analyses:}} In Figure \ref{fig: dice-vs-age}, as the age interval in-between moving/fixed pairs increases, the Dice score decreases for all three explored \rev{initialization approaches for both DL Reg and SyN ANTs}. Indeed, comparing how much of explained Dice score variance is due to the age interval when averaging over segmented regions for DL \rev{R}eg versus SyN ANTs using NR \revision{(67.0\% vs 50.8\%), RR (72.5\% vs 72.7\%) and RAR (72.4\% vs 72.3\%) initializations}, it shows similar results regardless of the approach. One could think this decrease in Dice score performance is due to the bigger proportion of examples with lower age intervals which the unsupervised network trains on (average age interval of 1.152±0.684, max age interval of 3.372 and min age interval of 0.114). However, the conventional iterative method which takes independent pairs also seems to be following the same negative trend between Dice scores versus age intervals negating this hypothesis. This potentially shows an intrinsic difficulty in registering brains which are quite different in age because of their topology, size and growth factors especially in the pediatric context. 
            
                \finalrev{\textbf{Sex-Specific Differences in Age-related Analyses:}} As for the supplementary Figures \ref{fig: noreg-sex} to \ref{fig: rigidaffinereg-sex}, showing Dice scores versus age intervals separated by sex (45\% female pairs and 54\% male pairs of total intra-subject pairs) for each initialization approach, the same trends seem to be followed where learning-based registration outperforms SyN ANTs, but both performances diminishing as the age interval increases. The exceptions are for NR, where \revision{for all three regions for males, DL Reg does not seem to vary with the age interval, whilst SyN ANTs is slightly more performant for small age intervals and decreases as the age interval increases. However, it's crucial to acknowledge that these differences are based on trendlines reflecting Dice scores, which exhibit significant variability due to the absence of prior initialization.} \revision{This trend shift is also visible for RR, but this time only for the CSF region only for males, where SyN ANTs is slightly more performant for age intervals bigger than 1.25 years}. The coefficients of correlation lie in the same range of around 72\% explained variance for all three initialization approaches versus SyN ANTs averaged over regions no matter the sex. \revision{The only exception is for NR, where the linear trend is much more visible for the DL Reg approach for males (average of 71.0\% explained variance) than females (61.8\%). In contrast, for SyN ANTs, the linear trend explains only 48.8\% for males and 53.4\% for females when averaged across all three segmented regions.}
                
                    \finalrev{\textbf{Recommendations:}} \revision{In light of the recommendations for pediatric longitudinal registration, it is crucial to acknowledge the trade-offs between speed, accuracy, and ease of use when selecting an appropriate \rev{registration} method. \rev{Both the RR + ANTs and RAR + ANTs methods provide rapid registration times, approximately 1.5 minutes per pair. Nevertheless, SyN ANTs frequently demands more individualized fine-tuning per pair compared to DL-based approaches, which, in contrast, leverage learned information from a specific population dataset all at once during the training process. It's important to note that the time spent on tuning SyN ANTs parameters is often not accounted for in provided times. Indeed, SyN ANTs exhibited lower performance as it wasn't precisely tuned for each region. For specific small structures like the caudate, amygdala, or pallidum, where precise registration accuracy is crucial, tweaking ANTs parameters might be relevant. On the other hand, if the objective involves studying multiple regions, both local and global, DL-based registration approaches offer a compelling alternative in terms of accuracy and efficiency when considering no fine-tuning is required when the network is trained for RR and RAR initializations.} Figure \rev{\ref{fig: dice-per-all-regions}} demonstrates that DL-based methods consistently slightly outperformed ANTs for almost all regions, as they were tailored for the specific population under consideration. \rev{It is also to be noted that there are no specific guidelines on how to pick SyN ANTs parameters which makes this task less reproducible. For younger cohorts, where the brain rapidly develops, a question that often arises is, should the parameters be optimized for each developmental stage or age range or standardized across stages?~\citep{Turesky2021-qv}. The answer depends on the dataset's characteristics, including available age ranges and the registration requirements, whether it involves segmenting all structures or only a smaller subset of local regions.}}
                    
                        \finalrev{\textbf{Limitations:}} Finally, it is worth noting that \revision{in comparison to SyN ANTs, learning-based algorithms exhibit a notable reliance on their training set, while SyN ANTs proves effective with any intra-subject image pair. \rev{The average pro-rated training time for DL Reg, obtained by dividing the total training time by the number of training pairs, is approximately 127.51±2.37 across all initialization approaches. In contrast, SyN ANTs has a zero-second training time since registration is conducted individually for each pair and no learning process is involved. However, achieving optimal registration alignments in terms of Dice scores requires manual tweaking of the SyN parameters. Despite longer inference times, SyN ANTs demonstrates greater generalizability. It is essential to acknowledge the generalization limitations of learning-based algorithms, as the trained network exhibits faster inference times specifically for the chosen pediatric dataset within the provided age range of 2-7 years old subjects. Assessing these strategies on new pediatric MRI datasets within the same age range would shed light on their adaptability to entirely unseen data. While 5-fold cross-validation offers insights into generalization capabilities, testing on a new dataset would provide a more robust assessment.} However, given the challenges of obtaining multiple longitudinal time-points in pediatric datasets\rev{~\citep{Wang2023-kr}}, especially within specific age ranges with limited data availability, testing these learning-based strategies on new data becomes notably challenging.}
                    
                        After comparing both DL Reg and SyN ANTs \rev{for all three initialization approaches}, it is \rev{noteworthy} that \rev{DL-based approaches with RR and RAR initializations}, show promising results, delivering Dice scores comparable to SyN ANTs but at significantly faster \rev{inference times. However, ANTs' advantage is that it does not require training.} RR and RAR excel in registering intra-subject images, particularly due to RAR's robust pre-alignment strategy. On the other hand, NR encounters challenges in capturing both local and global transformations. Age-related analyses reveal a consistent trend of decreasing Dice scores with larger age intervals, a phenomenon observed across all methods. \revision{There are some sex-specific shifts in performance noted for NR in all three tissues for males where SyN ANTs is slightly better than DL Reg for smaller age intervals.} The exact extent and significance of this influence would require further investigation and analysis. Finally, \revision{recommendations highlight the trade-offs between speed, accuracy, and ease of use, providing insights into the suitability of \rev{DL-based versus conventional SyN ANTs registration} methods for specific applications. The generalization capabilities of learning-based algorithms and the challenges of testing on new pediatric datasets are acknowledged.}

\section{Conclusion}
\finalrev{This study compared the conventional state-of-the-art SyN ANTs registration method with a DL-based approach, evaluating their performance in terms of accuracy, speed, initialization, and their influence on age intervals per pair within the intra-subject pediatric context.} Three initialization approaches were explored: NR (without initialization), RR (rigid initialization) and RAR (rigid and affine initialization). \finalrev{Registration quality was evaluated using both unweighted and weighted Dice scores for WM, GM, and CSF segmentations, averaged from corresponding sub-regions of the 18 available regions, while also assessing the individual performance of each region.} \revision{Additionally, we computed the average sum of absolute log JD values to assess the regularity of the obtained deformation fields for both DL Reg and SyN ANTs methods. These two methods showed comparable regularization strengths for both within \rev{the brain region} and per segmented region. We demonstrate that learning-based approaches, both with linear pre-alignments (RR and RAR) and without (NR), exhibit slight superiority over the SyN ANTs registration method in terms of Dice scores. Regarding registration quality, the DL-based approaches, specifically with RR and RAR initializations, significantly outperform (p $<$ 0.05) SyN ANTs, showing mean Dice scores across all 18 regions of 0.825±0.024 versus 0.818±0.017 and 0.913±0.018 versus 0.895±0.023, respectively. In RR and RAR initializations, SyN ANTs often falls short of surpassing Dice scores achieved during the initial alignment, especially in RAR, indicating challenges in detecting local changes related to neurodevelopment.} \finalrev{The main takeaway is that DL-based methods offer faster and more accurate registrations, while SyN ANTs is robust and works well without needing extensive training. Choosing the right method depends on the registration scale needed. DL-based registration adapts quickly but needs training, which can take time, while SyN ANTs requires manual adjustments and lacks clear guidelines, particularly regarding parameter choices for younger cohorts.} Both conventional and unsupervised DL-based approaches had their Dice scores decrease as the age interval increased showing intrinsic difficulties to register greater growth changes. Hence, faster registration steps of pairs closer in time can be used to uncover growth characterizations for future pediatric neurodevelopmental pipelines.

    Future work will apply this framework on younger populations (0-2 years) \rev{from other datasets} where the developmental factor is greater. These changes can be inspected in terms of their obtained DDFs to uncover growth patterns in the intra-subject context. It would also be useful to decompose the network into both global and local elements to be able to more accurately register fine-grained and more local brain regions.

\acks{This study was supported by Polytechnique Montréal, by the Canada First Research Excellence Fund, by CHU Sainte-Justine Research Center and by the TransMedTech Institute.}

% Ethical Standards.
% Please edit with the appropriate ethics considerations for your work. Include any pertinent IRB information, etc.
%
% Please note that the submission requirements included:
% The work presented must follow appropriate ethical standards in conducting research and writing the manuscript, following all applicable laws and regulations regarding treatment of animals or human subjects.
\ethics{The work follows appropriate ethical standards in conducting research and writing the manuscript, following all applicable laws and regulations regarding treatment of animals or human subjects.}

% Conflict of Interest
% Declaration of possible conflicts of interest: Authors must disclose any financial, organisational, commercial or personal conflicts of interest that might bias their work.
% If no conflicts, please say "We declare we don't have conflicts of interest."
\coi{We declare we don't have conflicts of interest.}

\bibliography{sample}

% needed in general before appendices.
% \newpage

% Appendix is optional
\clearpage
\appendix
\section{Supplementary Material}

\subsection{Data Preprocessing} \phantomsection \label{supp-material}
\finalrev{Prior to using DL models, preprocessing of MR images is crucial. This involves N4 bias correction for intensity uniformity and skull-stripping for isolating the brain region. \rev{Skull-stripping was done by first doing  rigid, affine and SyN registration steps to the Montreal’s Neurological Institute (MNI) 4.5-8.5 template~\citep{Fonov2011-jg} to obtain a brain mask for each subject scan. For rigid and the affine transforms, the parameters used are a gradient of 0.1, mattes similarity metric, 1000x11110x11110 multi-resolution steps, a threshold of 1e-7 for 20 iterations as a convergence criteria, 3x2x1 shrink factors and 4x2x1 voxels as smoothing sigmas. As for SyN, its gradient step, updateFieldVarianceInVoxelSpace and parameters are respectively 0.2, 3 and 0. A cross-correlation similarity metric is used with 100x100x50 multi-resolution steps, a threshold of -0.01 for 5 iterations as a convergence criteria, 4x2x1 shrink factors and 1x0.5x0 voxels as smoothing sigmas. The registrations performed were done at full resolution to ensure better alignment results. However, keeping the images at full resolution also increases registration time.} \cite{paniukov-2020}'s techniques, executed through Nipype in Python pipelines~\citep{Gorgolewski_et_al2021-fr}, were adopted for these tasks, \rev{as the analysis was successfully done on the same dataset. Moreover, a manual quality control (QC) step was incorporated to ensure accurate registration-based skull-stripping for each scan, and no pairs were excluded as a result. The final skull-stripped images were used as inputs to both SyN ANTs and the U-Net after rescaling them to 1.5x1.5x1.5 mm isotropic.}

As for the parameters used for rigid as well as rigid and affine initializations, they are a gradient of 0.1, mattes similarity metric, 500x250x100 multi-resolution steps, a threshold of 1e-6 for 10 iterations as a convergence criteria, 4x2x1 shrink factors and 2x1x0 voxels as smoothing sigmas. If some registrations failed with specific image pairs with the given parameters, the multi-resolution steps were increased, and the threshold decreased. \revision{For SyN ANTs, we employed a gradient step of 0.1, and updateFieldVarianceInVoxelSpace and totalFieldMeshSizeAtBaseLevel parameters were respectively set to 3 and 0.}}
	
\subsection{Metrics} \phantomsection \label{metrics}
\subsubsection{Unweighted and Weighted Dice Scores}
\finalrev{We use the term \textit{unweighted Dice score} to refer to the Dice score equation commonly used in registration studies, which assesses registration performance in terms of volume overlap between the warped image and ground truth. This metric is also interchangeably referred to as simply the Dice score throughout the article. These unweighted Dice scores are also inspected with respect to the age interval per pair and separated by sex. Dice scores are calculated in the white matter (WM), gray matter (GM) and cerebrospinal fluid (CSF) by averaging SynthSeg sub-regions within these tissues from the total 18 regions available. Given v, the voxels for the fixed (F) and moved volume (M), the Dice score for a region, r is calculated as follows:
\begin{align}
			Dice(v_\text{F}^\text{r},v_\text{M}^\text{r}\circ\phi)=2\cdot\frac{|v_\text{F}^\text{r}\cap(v_\text{M}^\text{r}\circ\phi)|}{|v_\text{F}^\text{r}|+|v_\text{M}^\text{r}\circ\phi|}
			\label{eq:dice}
		\end{align}

An additional analysis was introduced to reduce the influence of larger regions, preventing them from disproportionately elevating the mean Dice score across all 18 regions segmented using SynthSeg. Hence, a weighted Dice score, $D_{\mathrm{weighted}}$, is calculated per region, r, using a weight, $w_{\mathrm{r}}$ by incorporating the inverse number of voxels as presented in the equations \ref{eq:dice_weighted} to \ref{eq:factor} below. The average weighted Dice scores are obtained by summing the scores across all 18 segmented regions and subjects.
\begin{align}
			D_{\mathrm{weighted}}=\sum_{\mathrm{r}=1}^{18}w_{\mathrm{r}}\times D_{\mathrm{unweighted}}({\mathrm{r}})
			\label{eq:dice_weighted}
		\end{align}
$w_r$ is calculated as such per region where the denominator ensures that the weights collectively add up to 1 per subject,
\begin{align}
			w_{\mathrm{r}}=\frac{W_{\mathrm{r}}}{\sum_{\mathrm{r}=1}^{18}W_{\mathrm{r}}},
			\label{eq:weight}
		\end{align}
and the main weighting factor is the inverse of the number of voxels in that specific region as follows,
\begin{align}
			W_{\mathrm{r}}=\frac{1}{V_{\mathrm{r}}}
			\label{eq:factor}
		\end{align}}
\subsubsection{Negative JD and Sum of Absolute Log JD values}
\finalrev{We compute two key metrics related to the JD to assess deformation fields within local sub-regions extracted from SynthSeg and encompassing global white matter (WM), gray matter (GM), and cerebrospinal fluid (CSF) regions, excluding background regions. The JD quantifies volumetric changes following registration processes. 
The sum of absolute log JD values allows us to have a measure of the spread of the distribution of the JDs. For a volume of size $m$x$n$x$p$, it is calculated as represented in equation \ref{eq:JD} below. The average absolute sum of log Jacobian determinants is obtained by dividing the sum by the total number of voxels in the brain region per volume. 
\begin{align}
			\sum_{i=1}^m\sum_{j=1}^n\sum_{k=1}^p|\log(\text{JD}(i,j,k))|
			\label{eq:JD}
		\end{align}
As a negative JD signifies non-invertible properties, indicating the presence of undesired local folding, we also quantify the number of negative JDs using the following expression:
\begin{align}
			\text{\% of Negative JD values} = \frac{\sum_{i=1}^m\sum_{j=1}^n\sum_{k=1}^p1(\text{JD}(i,j,k)<0)}{m\cdot n\cdot p}\cdot100\
			\label{eq:JD2}
		\end{align}}
  % \max\{0,-JD(i,j,k)}
% These metrics offer insights into the deformation characteristics within specific brain regions and potential distortions, aiding in the assessment of the regularization strengths between methods.}
% \newpage
 % \begin{figure}[!htb]
	% \centering
	% \includegraphics[width=1\linewidth]{fig-a2.png}
	% \caption{Scheme showing the process used to obtain skull-stripped as well as WM, GM and CSF segmentations via independent subject by subject non-linear registration to the MNI 4.5-8.5 template~\citep{Fonov2011-jg}}
 %        \label{fig: skull-strip}
	% \end{figure}
 \newpage
\subsection{Representation of the Longitudinal Data}
\begin{figure}[!htb]
	\centering
	\includegraphics[width=0.5\linewidth]{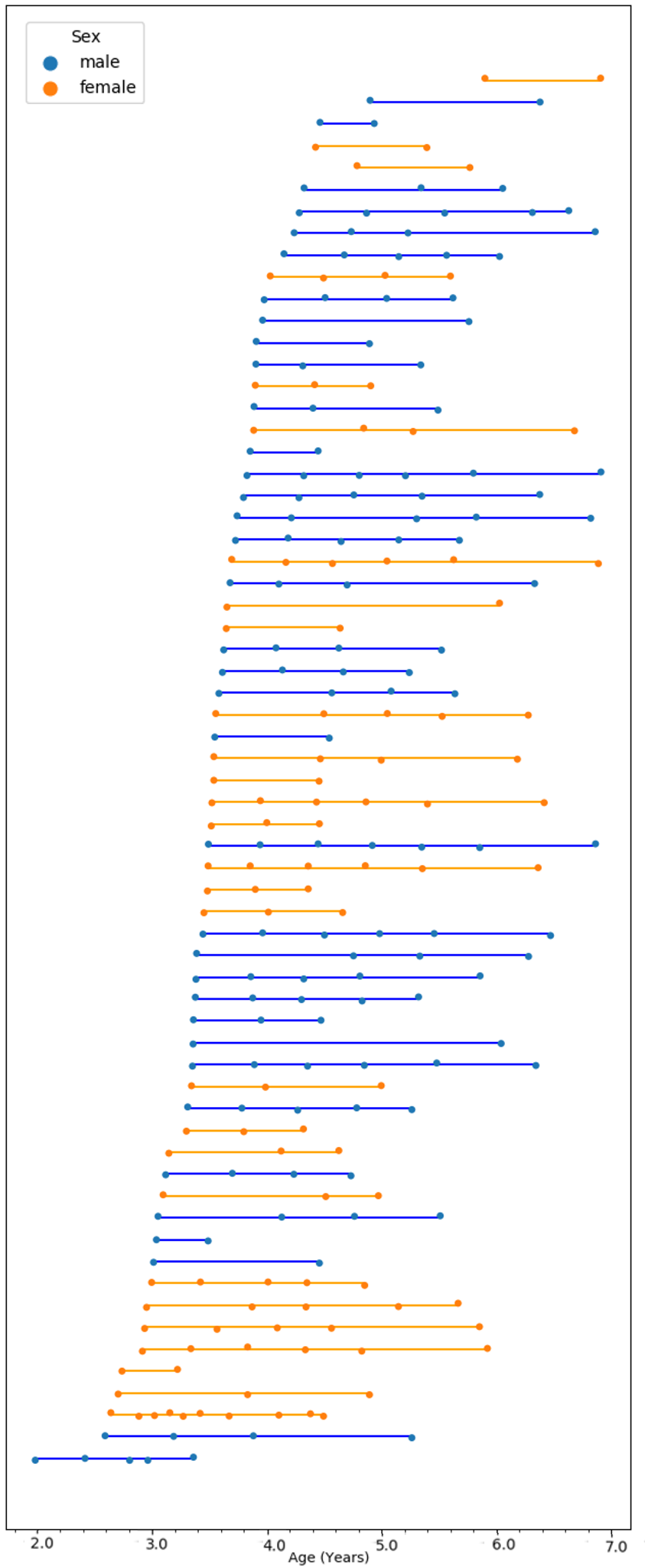}
	\caption{Representation of the 64 subjects used with age at scans and biological sex information, image inspired by~\citep{Reynolds524785}.}
        \label{fig: data-info}
	\end{figure}
\begin{landscape}
 \subsection{Full Pipeline}
 \begin{figure}[!htb]
	\centering
	\includegraphics[width=1\linewidth]{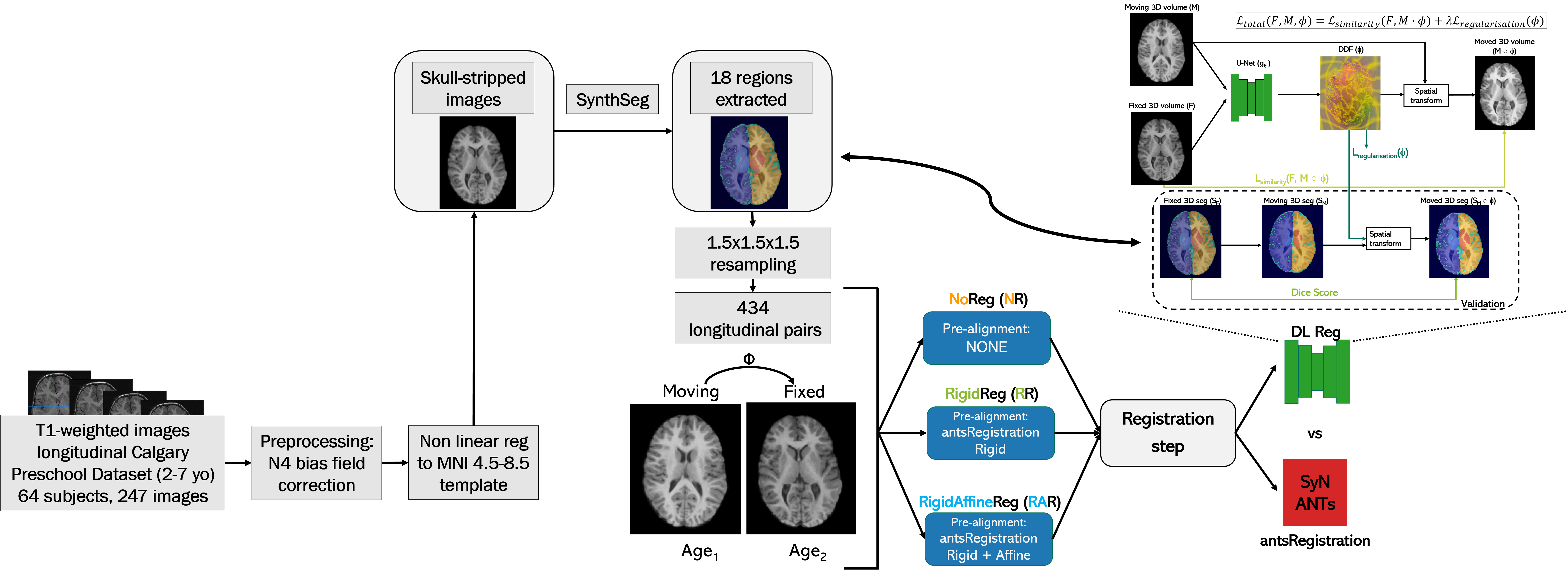}
	\caption{\revision{Full pipeline: 247 images from the longitudinal Calgary Preschool dataset are first N4 bias field corrected then nonlinearly registered to the MNI 4.5-8.5 template to obtain skull-stripped. These skull-stripped images are segmented by DL-based $SynthSeg^+$~\citep{Billot2023-po}, a robust segmentation method, then resampled to 1.5 mm isotropic resolution. All longitudinal pairs per subject (average: 3.86±1.59 time points/subject) are pre-aligned considering three initialization approaches (NoReg, RigidReg and RigidAffineReg). Deformations obtained by a unsupervised registration scheme using DeepReg are compared to the conventional SyN ANTs method using Dice scores.}}
        \label{fig: full-pipeline}
	\end{figure}
\end{landscape}

 %  \begin{figure}[!htb]
	% \centering
	% \includegraphics[width=1\linewidth]{fig-a4.png}
	% \caption{GroupKFold from scikit-learn~\citep{Pedregosa2011-yz} is applied to 64 subjects using 5 folds to never test with the same subjects present in the training set.}
 %        \label{fig: k-fold}
	% \end{figure}
\subsection{Weighted Dice Scores Across All 18 Regions}
 \begin{figure}[!htb]
	\centering
	\includegraphics[width=1\linewidth]{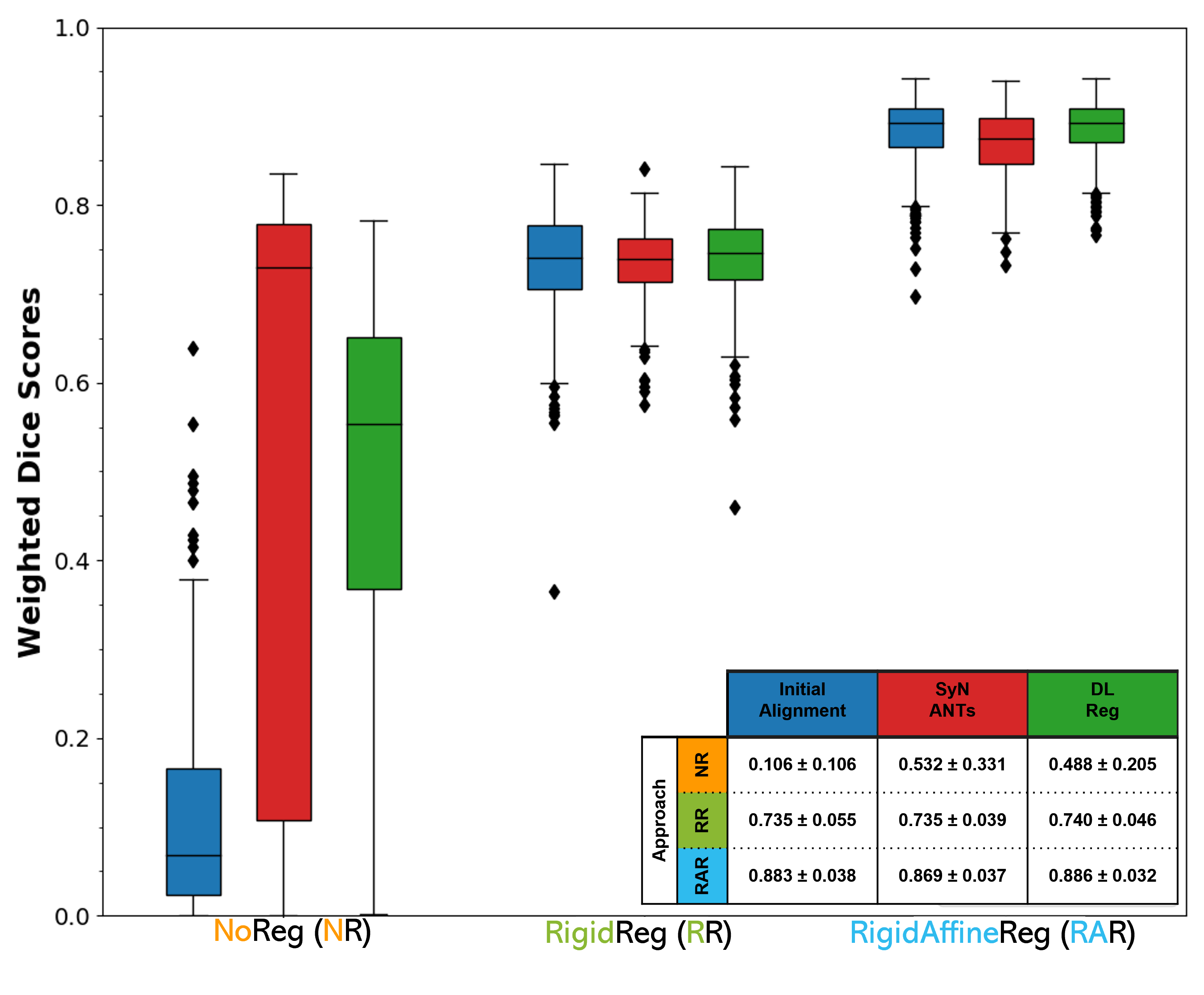}
	\caption{\rev{Weighted Dice score results on the test sets represented as boxplots for each initialization approach (NoReg, RigidReg and RigidAffineReg) for DL-based methods compared to the initial Dice scores pre-conducting the registration steps. Each method is also compared to the SyN ANTs registration. The weighting is determined by the inverse number of voxels per region, normalized by the sum of weights for all 18 regions, ensuring they collectively add up to 1 per subject. The average weighted Dice scores are computed by summing across all 18 segmented regions and subjects, considering their respective weights. The table in the lower right corner provides the mean±SD Dice scores for all scenarios.}}
        \label{fig: dice-results-weighted}
	\end{figure}
 \newpage
\subsection{Dice Scores in Global Tissues}
 \begin{figure}[!htb]
	\centering
	\includegraphics[width=1\linewidth]{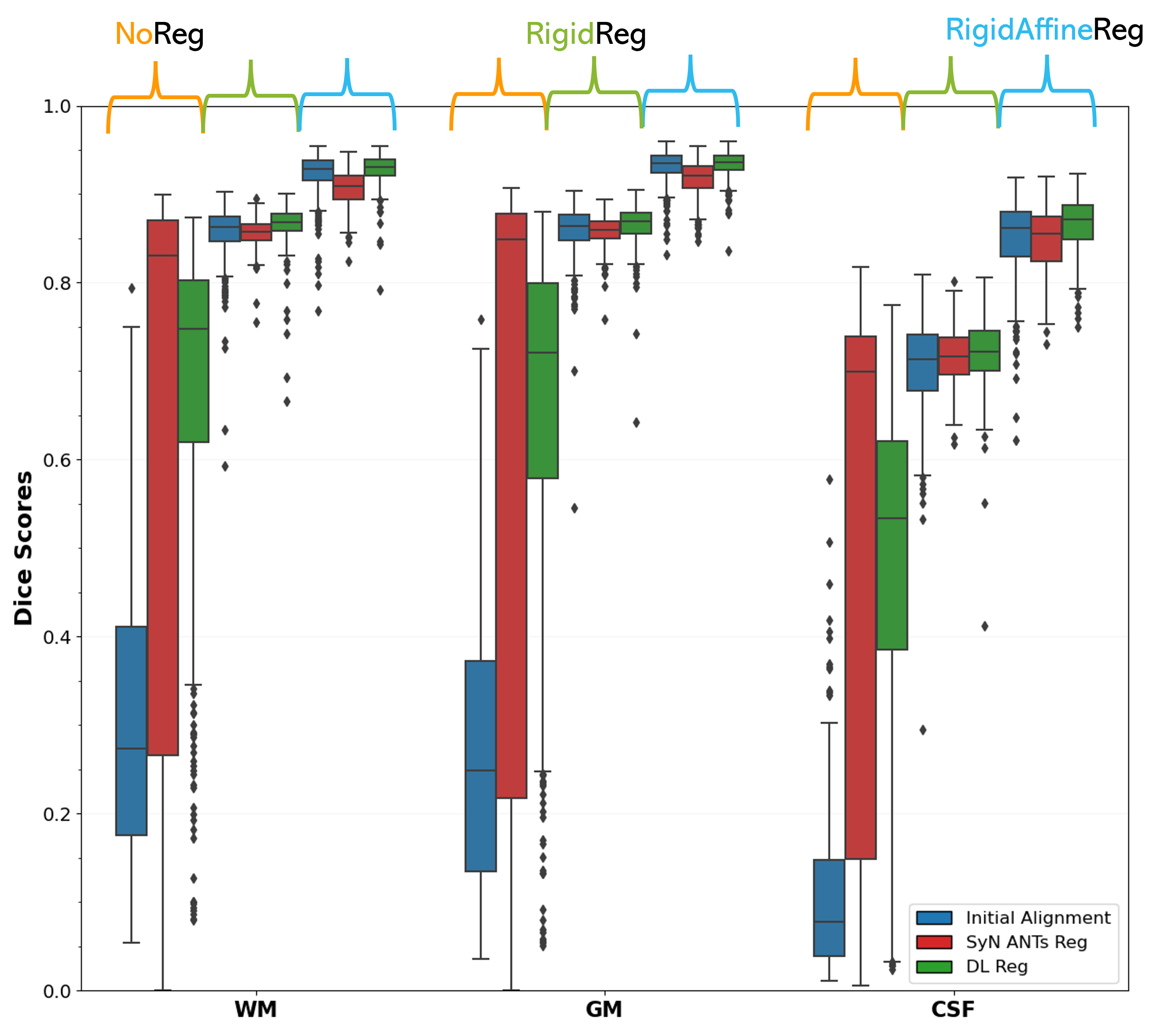}
	\caption{\revision{Dice score results on the evaluation set represented as boxplots for each method (NoReg, RigidReg and RigidAffineReg indicated as respectively orange, light green or light blue braces) compared to the initial \rev{Dice} scores pre-conducting the registration steps. Each method is compared to SyN ANTs registration (refer to Figure \ref{fig: method-scheme}). Dice scores are computed for the white matter (WM), gray matter (GM), and cerebrospinal fluid (CSF) by averaging SynthSeg sub-regions within these tissues from the total 18 regions available.}}
        \label{fig: dice-per-region}
	\end{figure}
 \subsection{Dice Scores per Region}
 \begin{landscape}
 % \vspace*{-40mm}
 % \filbreak
 % \subsection{Dice Scores per Region}
  % \vspace*{-100mm}
  \begin{figure}[!htb]
	\centering
	\includegraphics[width=1\linewidth]{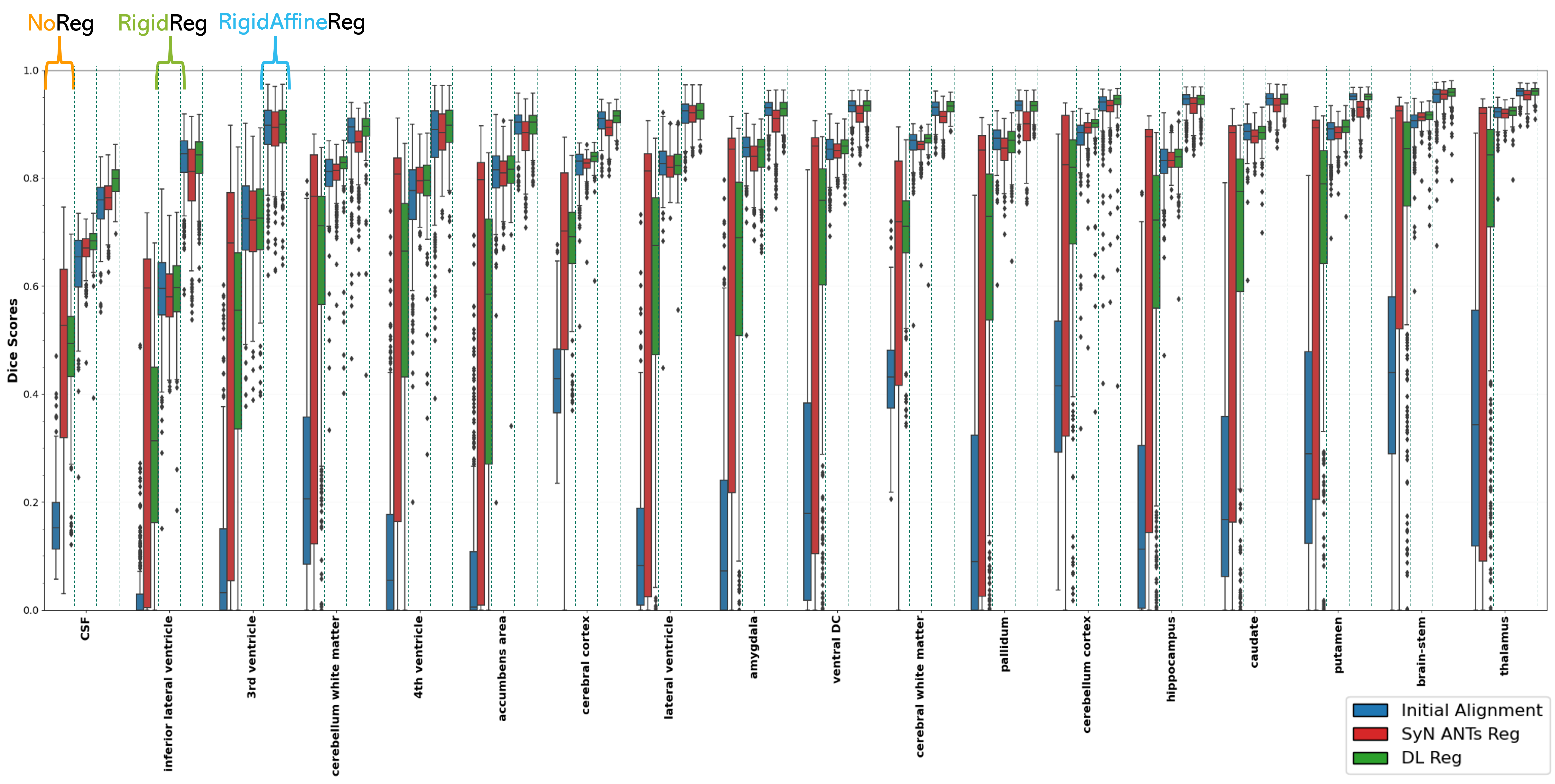}
	\caption{\revision{Boxplots depict Dice score results on the evaluation set for each method (NoReg, RigidReg, and RigidAffineReg denoted by orange, light green, or light blue braces, respectively) in comparison to the initial \rev{Dice} scores before executing the registration steps. Each method is contrasted with SyN ANTs registration (refer to Figure \ref{fig: method-scheme}). Dice scores are computed \rev{for every region amongst the 18 regions extracted from SynthSeg}, averaging left and right brain hemisphere labels for all structures except brain stem and CSF. Regions are arranged \rev{in ascending order} according to the Dice scores obtained with DL-based RigidAffineReg.} \rev{For reference, the 18 brain regions, ordered from smallest to largest by the number of voxels, are as follows: 3rd ventricle, inferior lateral ventricle, accumbens area, 4th ventricle, amygdala, pallidum, ventral DC, hippocampus, caudate, lateral ventricle, putamen, thalamus, brain-stem, cerebellum white matter, cerebellum cortex, CSF, cerebral white matter, and cerebral cortex.}}
        \label{fig: dice-per-all-regions}
	\end{figure}

 \end{landscape}
 \subsection{Visualisations of Example Pairs}
 \begin{figure}[!htb]
	\centering
	\includegraphics[width=1\linewidth]{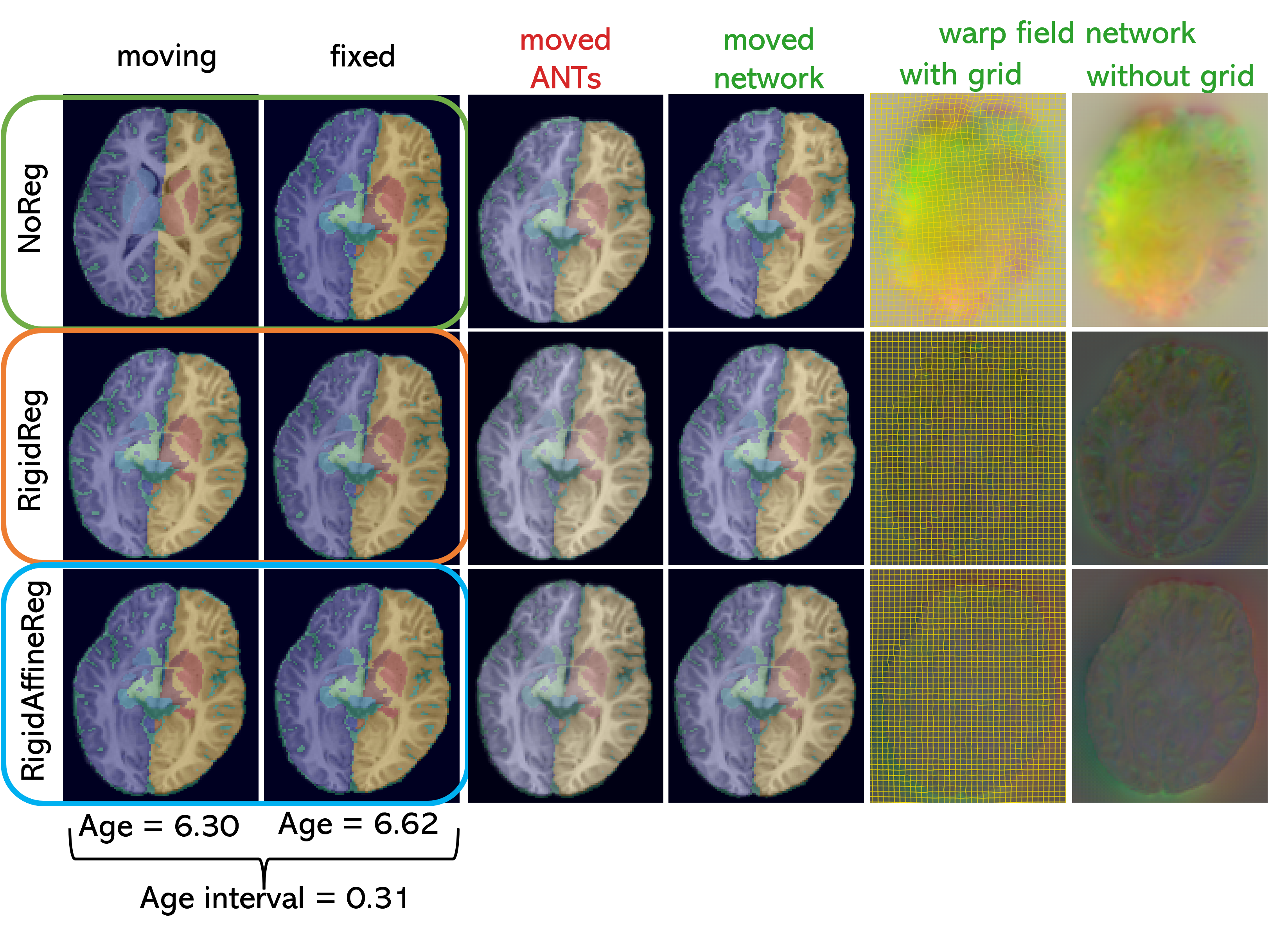}
	\caption{\revision{Illustration of overlaid moving, fixed, and transformed segmentations on the intensity volumes for each of the three initialization methods (NoReg, RigidReg and RigidAffineReg) and both ANTs and DL-based pipelines for a subject with an age interval of 0.31. The fourth and fifth columns depict the warping grid overlaid with the RGB image of displacement values in each spatial dimension and the RGB image itself, respectively.}}
        \label{fig: example-fields}
	\end{figure}
 \newpage
\subsection{P-Values and Statistical Significance Analysis}
\begin{figure}[!htb]
	\centering
	\includegraphics[width=1\linewidth]{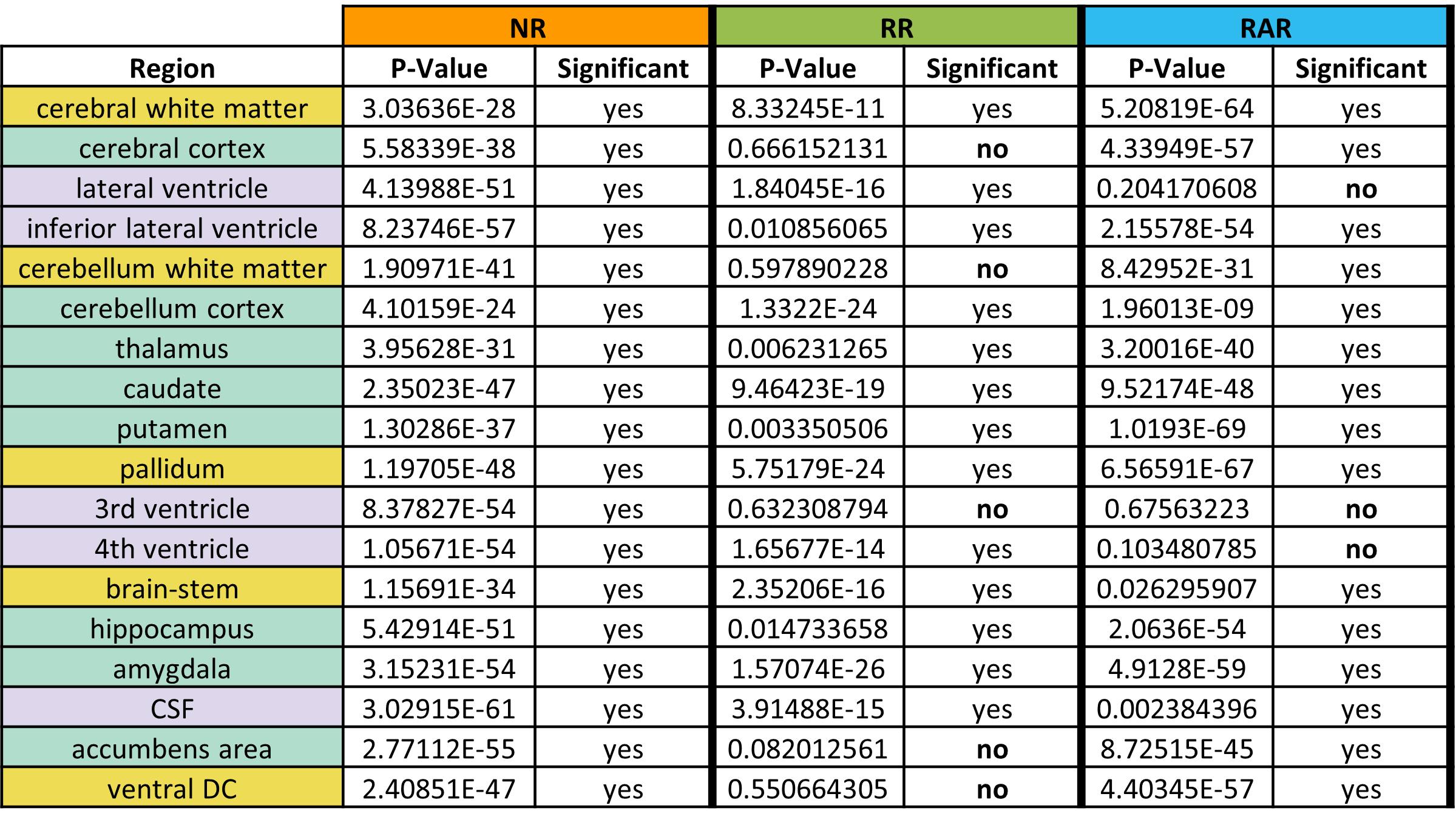}
	\caption{\finalrev{P-values and statistical significance (p$<$0.05) derived from a Wilcoxon test comparing SyN ANTs Dice scores to initial Dice scores before any initialization, across all 18 individual regions. The colors in the table correspond to global regions, with white matter in yellow, gray matter in turquoise, and cerebrospinal fluid in purple, matching the color scheme outlined in Table \ref{fig: table-final}. Each set of two columns represents data from one of the three initialization approaches (NoReg, RigidReg, and RigidAffineReg).}}
        \label{fig: ANTs_init_stats}
	\end{figure}
\newpage
 \begin{figure}[!htb]
	\centering
	\includegraphics[width=1\linewidth]{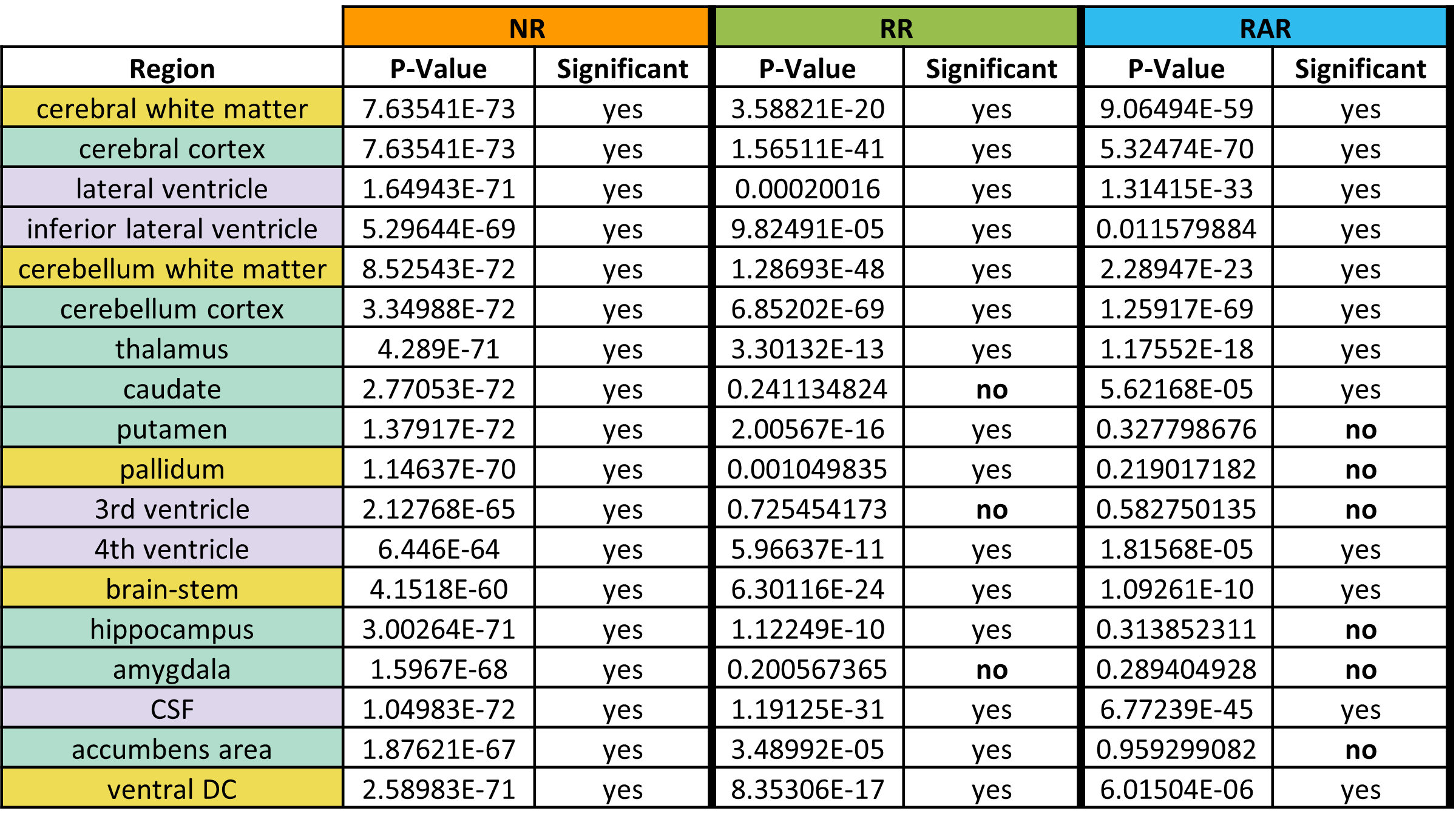}
	\caption{\finalrev{P-values and statistical significance (p$<$0.05) derived from a Wilcoxon test comparing DL Reg Dice scores to initial Dice scores before any initialization, across all 18 individual regions. The colors in the table correspond to global regions, with white matter in yellow, gray matter in turquoise, and cerebrospinal fluid in purple, matching the color scheme outlined in Table \ref{fig: table-final}. Each set of two columns represents data from one of the three initialization approaches (NoReg, RigidReg, and RigidAffineReg).}}
        \label{fig: DL_init_stats}
	\end{figure}
\newpage
 \begin{figure}[!htb]
	\centering
	\includegraphics[width=1\linewidth]{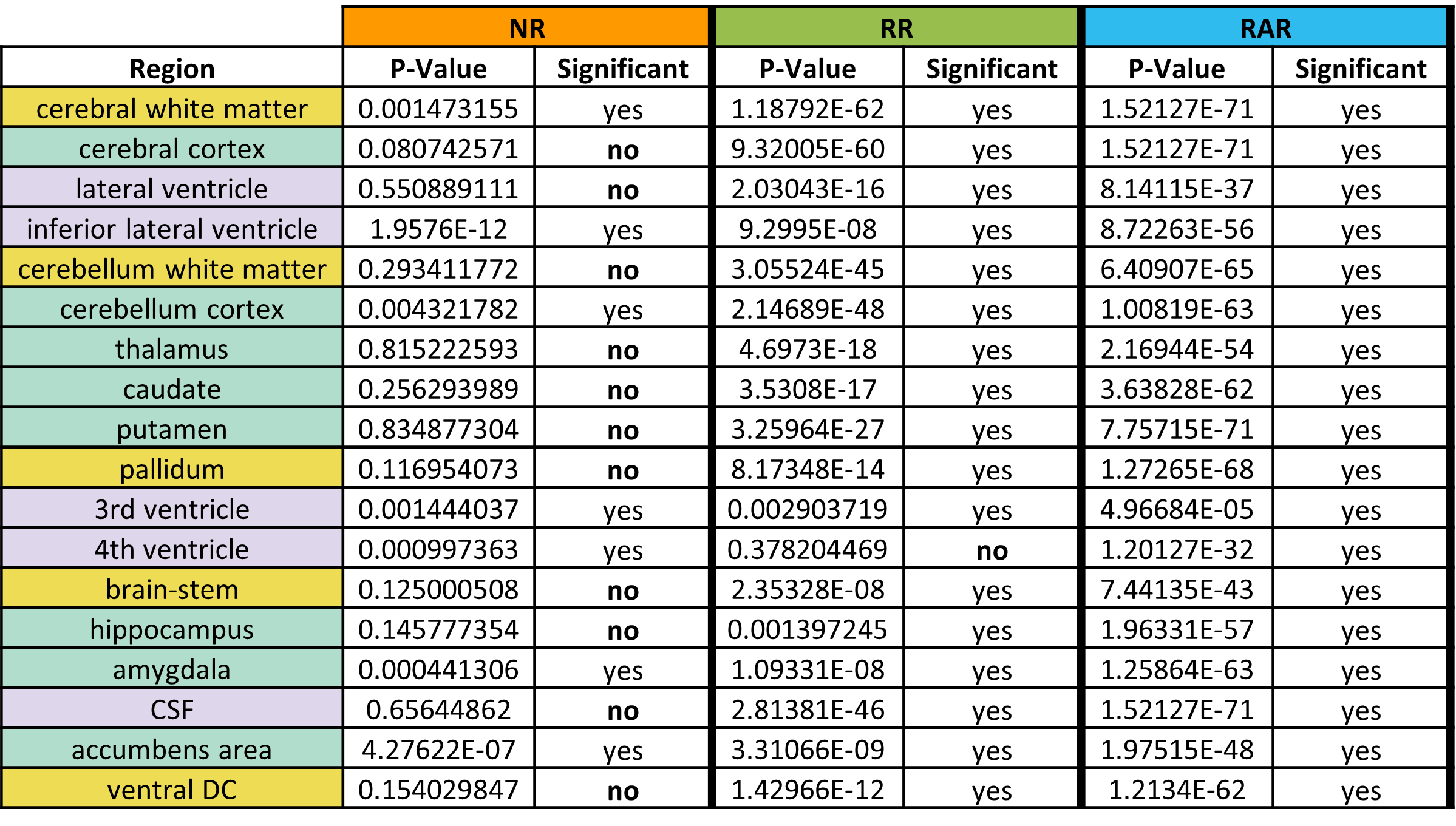}
	\caption{\finalrev{P-values and statistical significance (p$<$0.05) derived from a Wilcoxon test comparing DL Reg Dice scores to SyN ANTs Dice scores, across all 18 individual regions. The colors in the table correspond to global regions, with white matter in yellow, gray matter in turquoise, and cerebrospinal fluid in purple, matching the color scheme outlined in Table 2. Each set of two columns represents data from one of the three initialization approaches (NoReg, RigidReg, and RigidAffineReg).}}
        \label{fig: DL_ANTs_stats}
	\end{figure}

 %   \begin{figure}[!htb]
	% \centering
	% \includegraphics[width=1\linewidth]{fig-a11.png}
	% \caption{Sum of absolute Log Jacobian determinants for all three segmented regions compared with regards to DL-based registration (DL Reg) versus SyN ANTs for each initialization (NoReg, RigidReg and RigidAffineReg indicated as respectively orange, light green or light blue braces).}
 %        \label{fig: sum-jacobians-per-region}
	% \end{figure}
 \newpage
\subsection{Sex-Specific Age-Related Analyses}
  \begin{figure}[!htb]
	\centering
	\includegraphics[width=1\linewidth]{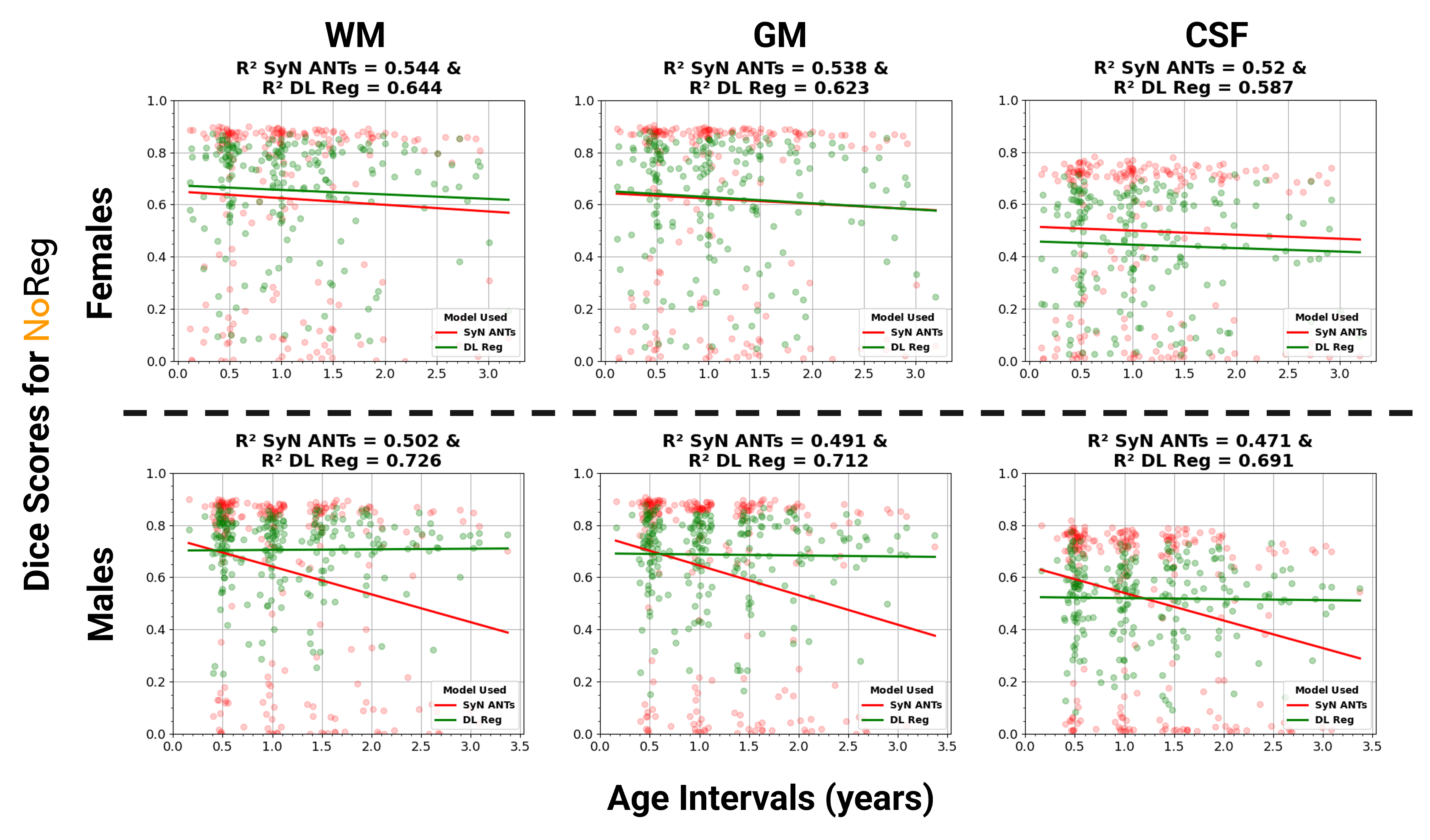}
	\caption{Dice score results compared to the age interval between moving and fixed pairs calculated on the test set separated by sex, females on the first row and males on the second for no prior initialization and DL-based registration \revision{(DL Reg)} in green compared to SyN ANTs registration in red. The \rev{Dice} scores are calculated in the white matter (WM), gray matter (GM) and cerebrospinal fluid (CSF) \rev{by averaging SynthSeg sub-regions within these tissues from the total 18 regions available and are} presented on each column. Coefficients of determination (R-squared) are also presented in each figure title.}
        \label{fig: noreg-sex}
	\end{figure}

  \begin{figure}[!htb]
	\centering
	\includegraphics[width=1\linewidth]{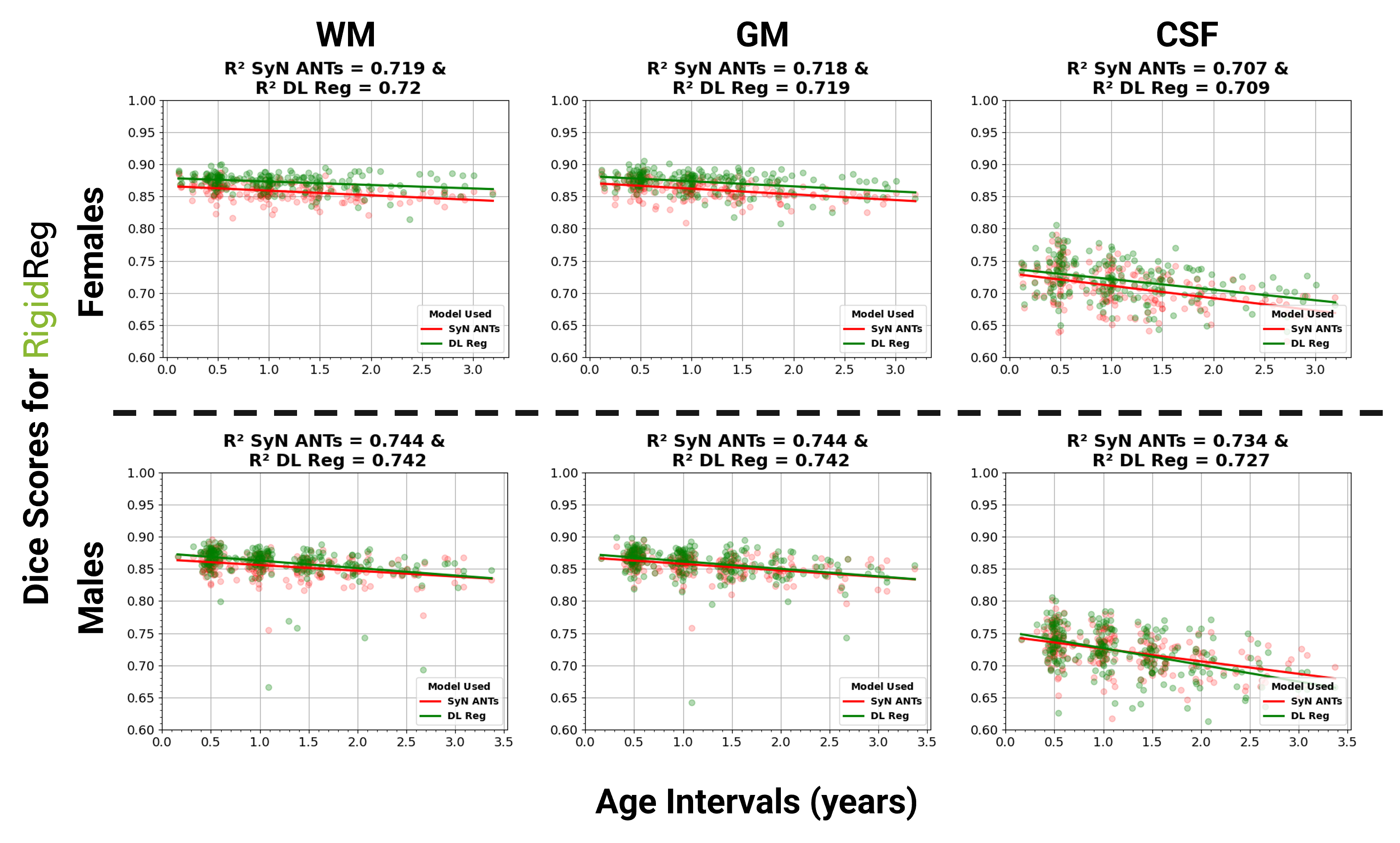}
	\caption{Dice score results compared to the age interval between moving and fixed pairs calculated on the test set separated by sex, females on the first row and males on the second for rigid initialization and DL-based registration \revision{(DL Reg)} in green compared to SyN ANTs registration in red. The \rev{Dice} scores are calculated in the white matter (WM), gray matter (GM) and cerebrospinal fluid (CSF) \rev{by averaging SynthSeg sub-regions within these tissues from the total 18 regions available and are} presented on each column. Coefficients of determination (R-squared) are also presented in each figure title.}
        \label{fig: rigidreg-sex}
	\end{figure}

 \begin{figure}[!htb]
	\centering
	\includegraphics[width=1\linewidth]{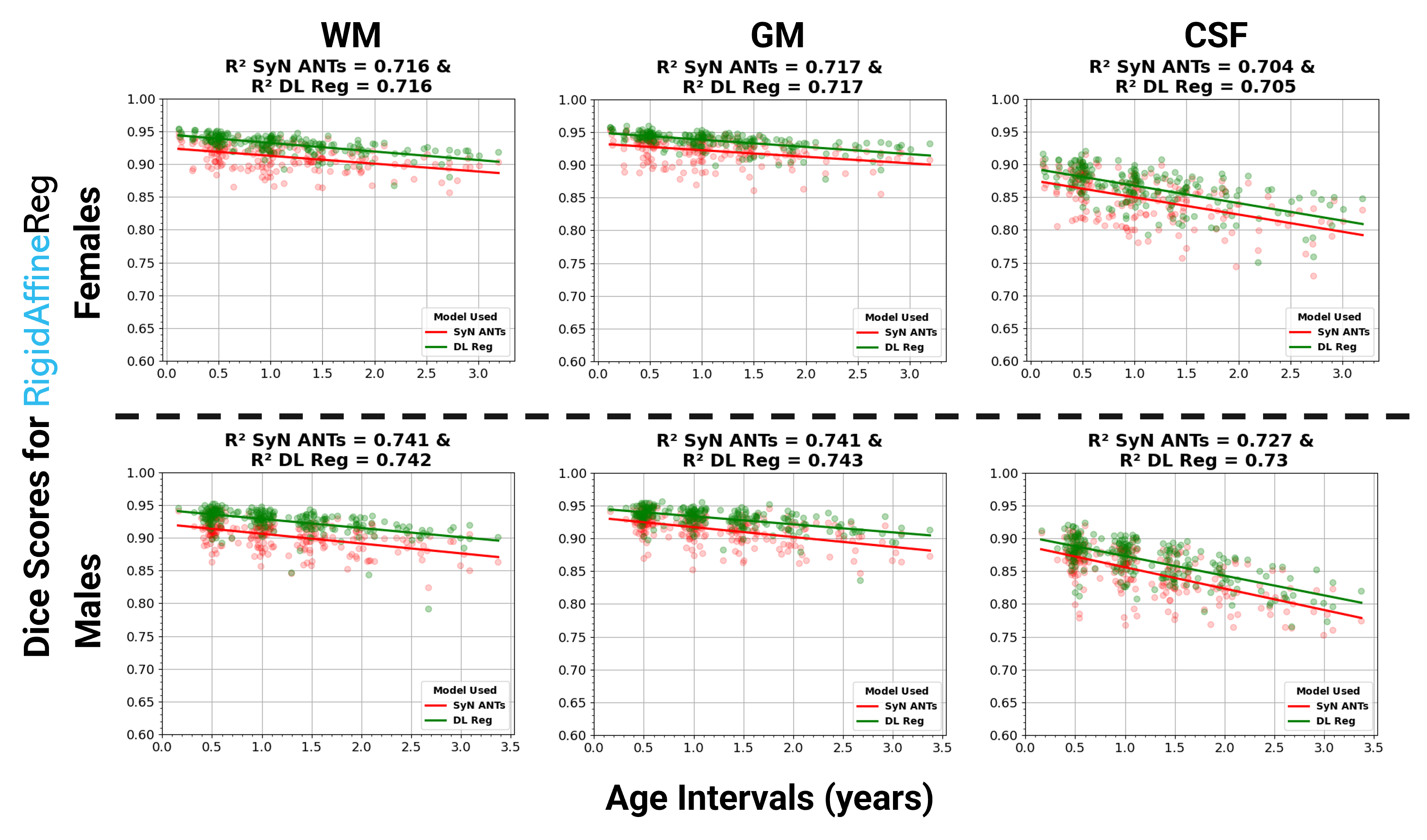}
	\caption{Dice score results compared to the age interval between moving and fixed pairs calculated on the test set separated by sex, females on the first row and males on the second for rigid and affine initialization and DL-based registration \revision{(DL Reg)} in green compared to SyN ANTs registration in red. The \rev{Dice} scores are calculated in the white matter (WM), gray matter (GM) and cerebrospinal fluid (CSF) \rev{by averaging SynthSeg sub-regions within these tissues from the total 18 regions available and are} presented on each column. Coefficients of determination (R-squared) are also presented in each figure title.}
        \label{fig: rigidaffinereg-sex}
	\end{figure}

	% {\noindent \em Remainder omitted in this sample. }

% \section{Appendix section}
% 	\subsection{Appendix subsection}
% 		\subsubsection{Appendix subsubsection}
% 			\paragraph{Appendix paragraph} Lorem ipsum dolor sit amet, consectetur adipisicing elit, sed do eiusmod
% 			tempor incididunt ut labore et dolore magna aliqua. Ut enim ad minim veniam,
% 			quis nostrud exercitation ullamco laboris nisi ut aliquip ex ea commodo
% 			consequat. Duis aute irure dolor in reprehenderit in voluptate velit esse
% 			cillum dolore eu fugiat nulla pariatur. Excepteur sint occaecat cupidatat non
% 			proident, sunt in culpa qui officia deserunt mollit anim id est laborum.

\end{document}